\documentclass[prd,nofootinbib,floats,superscriptaddress,eqsecnum,tightenlines,11pt]{revtex4}

\usepackage{hyperref}
\usepackage{graphicx}
\usepackage{amsmath,amssymb,amsfonts,amsthm,latexsym,stmaryrd}
\usepackage{marginnote}
\usepackage{color}

\usepackage{physics}
\usepackage{csquotes}
\usepackage{tensor}
\usepackage{mathtools}
\usepackage{amssymb}
\usepackage{siunitx}
\usepackage{caption}
\usepackage{cleveref}
\usepackage{multirow}
\usepackage{rotating}
\usepackage[toc,page]{appendix}

\usepackage{tikz}
\usetikzlibrary{shapes.geometric, arrows, positioning, matrix, shapes.misc, calc}

\def\be{\begin{equation}}
\def\ee{\end{equation}}
\def\beq{\begin{equation}}
\def\eeq{\end{equation}}
\newcommand{\bea}{\begin{eqnarray}}
\newcommand{\eea}{\end{eqnarray}}
\def\bi{\begin{itemize}}
\def\ei{\end{itemize}}
\def\ba{\begin{array}}
\def\ea{\end{array}}
\def\bfig{\begin{figure}}
\def\efig{\end{figure}}

\DeclareMathOperator{\cotan}{cotan}

\newcommand\mass{r_s}

\newcommand{\X}{Y}
\newcommand{\p}{r} 
\newcommand{\q}{f} 
\newcommand{\Id}{I} 

\tikzset{
	startstop/.style={
		rectangle, text width=3cm, minimum height=1cm, text centered, draw=black, fill=red!30, anchor=center
	},
	process/.style ={
		rectangle, text width=3cm, minimum height=1cm, text centered, draw=black, fill=orange!30, anchor=center
	},
	decision/.style ={
		chamfered rectangle, chamfered rectangle xsep=2cm, text width=3cm, minimum height=1cm, text centered, draw=black, fill=green!30, anchor=center
	},
	arrow/.style = {
		thick,->,>=stealth
	},
}

\definecolor{color1}{RGB}{240, 70, 135}
\definecolor{color2}{RGB}{49, 173, 38}
\definecolor{color3}{RGB}{46, 204, 240}

\begin{document}

\title{Asymptotics of linear differential systems and application to quasi-normal modes of  nonrotating black holes}

\author{David Langlois}
\affiliation{Universit\'e de Paris, CNRS, Astroparticule et Cosmologie, F-75006 Paris, France}
\author{Karim Noui}
\affiliation{Institut Denis Poisson (UMR 7013), Universit\'e de Tours, Universit\'e d'Orl\'eans, Parc de Grandmont, 37200 Tours, France}
\affiliation{Universit\'e de Paris, CNRS, Astroparticule et Cosmologie, F-75006 Paris, France}
\author{Hugo Roussille}
\affiliation{Universit\'e de Paris, CNRS, Astroparticule et Cosmologie, F-75006 Paris, France}
\affiliation{Institut Denis Poisson (UMR 7013), Universit\'e de Tours, Universit\'e d'Orl\'eans, Parc de Grandmont, 37200 Tours, France}

\date{\today}

\begin{abstract}
	The traditional approach to perturbations of  nonrotating black holes in General Relativity uses the reformulation of the equations of motion into a radial second-order Schr\"odinger-like equation, whose  asymptotic solutions are elementary. Imposing specific  boundary conditions at spatial infinity and near  the horizon defines, in particular, the quasi-normal modes of black holes.  For more complicated equations of motion,  as encountered for instance in modified gravity models with different background solutions and/or additional degrees of freedom, we present a new approach that analyses directly the first-order differential system  in its original form and extracts   the asymptotic behaviour of perturbations, without resorting to a second-order reformulation.  As a pedagogical illustration, we apply this treatment to the perturbations of Schwarzschild black holes and then show that the standard quasi-normal modes can be obtained numerically by solving  this first-order system with a spectral method. This new approach paves the way for a generic treatment of the asymptotic behaviour of black hole perturbations and the identification of quasi-normal modes in theories of modified gravity.
	
\end{abstract}

\maketitle

\section{Introduction}

The oscillations of black holes (BH) have been studied theoretically for several decades. Today, with the first observations of gravitational waves emitted by BH mergers, one can now hope to observe directly these oscillations via their GW signatures, especially in the ringdown phase of the signal when the post-merger black hole relaxes to a Kerr black hole, according to General Relativity. One of the major goals of future detections will be  to check whether the observed oscillations coincide with the predictions based on General Relativity 
(see e.g. \cite{Berti:2005ys,Berti:2018vdi}). 
This is also an ideal playground to test alternative theories of gravitation. Indeed, even if the background BH solution may coincide with that of GR, the linear perturbations  in general obey different equations of motion.

During the ringdown phase, at least in the linear regime, the GW signal is expected to mainly consist of a superposition of the so-called quasi-normal, or resonant, modes (QNMs)  which have been excited by the merger and then decay via GW radiation: these modes correspond to the proper oscillation modes of the black hole and are characterised by a complex frequency $\omega$, whose  imaginary part quantifies their damping rate.

 In the simplest case of nonrotating black holes, i.e. Schwarzschild black holes, the computation of QNMs  is based  on the  classical papers by Regge \& Wheeler \cite{Regge:1957td} and later Zerilli \cite{Zerilli:1970se}, who reformulated the linearised Einstein equations in the frequency domain, which are {\it first-order} with respect to the radial coordinate, as a {\it second-order}  Schr\"odinger-like equation. This familiar equation, with a specific potential for axial and polar metric perturbations, is the standard starting point for the numerical calculations   or semi-analytical treatments  of QNMs, using for instance well-known methods in quantum mechanics.

Understanding the asymptotic behaviour of the perturbations at the horizon and at spatial infinity is crucial for QNMs, which are defined by very specific  boundary conditions.  Indeed,  they 
correspond to purely outgoing radiation at spatial infinity and  ingoing radiation at the horizon. Imposing these specific boundary conditions leads to a discrete set of allowed frequencies. 

When the equations of motion of the perturbations are written as a second-order Schr\"odinger equation, obtaining their asymptotic behaviour is immediate, as it simply depends on   the asymptotic behaviour of the effective potential.  In the context of modified gravity however,  the problem can become more involved for several reasons. First, the background metric can differ from the standard GR solutions, i.e. be different from Schwarzschild in the nonrotating case. Moreover, modified theories often involve additional fields, such as scalar fields,  which increases the number of degrees of freedom and therefore the complexity of the linear equations of motion. 

In several interesting cases, the equations of motion can be rewritten as a generalised $N$-dimensional matrix Schr\"odinger-like system for $N$ fields $\Psi_i$, of the typical form (see e.g. \cite{McManus:2019ulj})
\beq
\label{schroedinger}
f \frac{d}{dr}\left(f \frac{d\Psi_i}{dr}\right)+\left(\omega^2-f \, V_{ij}\right) \Psi_j=0\,,
\eeq
where $f(r)=1-\mass/r$  and  the $N\times N$ matrix $V_{ij}$ of radial potentials  usually vanishes or becomes a constant diagonal matrix asymptotically. The frequency $\omega$ appears quadratically in the above system, which corresponds to a system of propagation equations if one replaces $\omega$ with $-i \pdv*{}{t}$. The boundary conditions are  still  easy to infer from such a differential system.

However, one could also encounter more general situations where such a simple reformulation of the equations of motion is not available or would require an involved and lengthy procedure. 
Specific examples will be given in a companion paper \cite{Langlois:2021aji},  in the context of 
  Degenerate Higher-Order Scalar-Tensor (DHOST) theories  \cite{Langlois:2015cwa,Crisostomi:2016czh,Achour:2016rkg,BenAchour:2016fzp} which provide the most general viable set of scalar-tensor theories to date.
  In those examples, it is not clear whether one can rewrite the polar equations of motion as a second-order Schr\"odinger-like system of the form \eqref{schroedinger}, with its specific dependence on  $\omega$. In the specific case of stealth Schwarzschild black holes, a lengthy manipulation of the quadratic Lagrangian for perturbations enabled the authors of \cite{Takahashi:2021bml}  to identify master variables, leading to a second-order differential system for the physical degrees of freedom, although of a more complex form than \eqref{schroedinger}.
To tackle more general situations,  it would be very useful to be able to analyse directly the first-order system of equations in its original form and to extract directly from it  the asymptotic behaviour of perturbations.

The purpose of this paper is to present  such a systematic treatment of a general first-order differential system.
In order to reach this goal, we use recent developments that appeared in the mathematical literature.  These results enable us to determine, via a systematic algorithm, the asymptotic structure of the solutions of a generic first-order differential system.  For pedagogical reasons, we use here  this algorithm to recover the asymptotic solutions for the axial, or odd-parity, modes and for the polar, or even-parity, modes of the standard Schwarzschild solution. This paper will be completed by a companion paper \cite{Langlois:2021aji} that applies the same method to a few black hole solutions in DHOST theories. 

The outline of the paper is the following. In the next section, we review the standard derivation for the Schwarzschild perturbations, distinguishing as usual the axial and polar modes. In section \ref{sec:Schwarzschild_first_order}, we present our new approach and show explicitly how this new method enables us to recover the usual asymptotic solution, working directly with the first order system. We also show how the quasi-normal modes can be computed in this new perspective. We then present, in section  \ref{sec:general_analysis}, the general algorithm,  carefully listing the various steps of the algorithm depending on the structure of the system. We give a summary and open some perspectives in the concluding section. A few appendices contain some additional details.

\section{A short review on Regge-Wheeler and Zerilli equations}
\label{sec1}

In this section, we review  the standard procedure to derive the equations of motion  for  the perturbations of a  Schwarzschild black hole in general relativity, originally obtained  by  Regge and Wheeler \cite{Regge:1957td} for the axial, or odd-parity, modes and Zerilli \cite{Zerilli:1970se} for the polar, or even-parity, modes.
These equations  can be shown to reduce to a Schr\"odinger-like equation with an effective potential characterising the ``dynamics'' of the linear perturbations. 

\subsection{Linear perturbations of Einstein equations about the Schwarzschild black hole}

We start with the four-dimensional Einstein-Hilbert action in vacuum (with no cosmological constant) for the metric $\tensor{g}{_\mu_\nu}$,
\begin{equation}
\label{eq:einstein-hilbert-action}
S[g_{\mu\nu}] = \frac{1}{16\pi G_N} \int \dd[4]{x} \sqrt{-g} \, R \, ,
\end{equation}
where $g \equiv \det(g_{\mu\nu})$ is the determinant of the metric, $R$ the four-dimensional Ricci scalar and $G_N$ denotes Newton's
constant,  which actually  will not show up in the equations of motion since we are not considering any  matter field here.

\subsubsection{Linearised general relativity}
Given any background  metric $\tensor{\overline{g}}{_\mu_\nu}$  solution to the Einstein equations, one can introduce the perturbed metric 
\beq
\tensor{g}{_\mu_\nu}=\tensor{\overline{g}}{_\mu_\nu}+ \tensor{h}{_\mu_\nu} 
\eeq
where the $\tensor{h}{_\mu_\nu}$ denote the linear  perturbations of the metric. In order to derive the linear equations of motion that govern the evolution of $h_{\mu\nu}$, one expands  the Einstein-Hilbert action \eqref{eq:einstein-hilbert-action} up to the second order in $ \tensor{h}{_\mu_\nu} $. The Euler-Lagrange equations associated with the {\it quadratic} part of this expansion then provide the linearised equations of motion for $h_{\mu\nu}$. 

By expanding \eqref{eq:einstein-hilbert-action}, one obtains  the following quadratic action for $h_{\mu\nu}$,
\bea
\label{eq:perturbed-EH-action}
S_{\rm quad}[h_{\mu\nu}] & = & \frac{1}{16\pi G_N} \int \dd[4]{x} \sqrt{-\overline{g}} \left\{
-\frac12 \tensor{h}{_\mu_\nu}\tensor{h}{^\mu^\nu} \bar R
+ \frac14 h^2 \bar  R
+ h \tensor{h}{_\mu_\nu} \tensor{\bar R}{^\mu^\nu}
+ 4 \tensor{h}{_\mu^\rho} \tensor{h}{^\mu^\nu} \tensor{\bar R}{_\nu_\rho} 
- 2 \tensor{h}{^\mu^\nu} \tensor{h}{^\rho^\sigma} \tensor{\bar R}{_\mu_\rho_\nu_\sigma}\right. \nonumber\\
&&\left. \hspace{1.5cm}+ \frac12 (\bar\nabla_\mu h) (\bar\nabla^\mu h)
- 2 (\bar\nabla_\mu \tensor{h}{^\mu_\nu}) (\bar\nabla_\rho \tensor{h}{_\nu^\rho})
- (\bar\nabla_\mu h)(\bar\nabla_\nu \tensor{h}{^\mu^\nu}) \right. \nonumber \\
&&\left.  \hspace{1.5cm} + 3 (\bar\nabla_\nu \tensor{h}{_\mu_\rho}) (\bar\nabla^\rho \tensor{h}{^\mu^\nu})
- \frac12 (\bar\nabla_\rho \tensor{h}{_\mu_\nu}) (\bar\nabla^\rho \tensor{h}{^\mu^\nu})
\right\},
\eea
where $\bar R_{\mu \nu \rho \sigma}$, $\bar R_{\mu\nu}$, $ \bar R$  and  $\bar\nabla_\mu$  are respectively the Riemann tensor, the Ricci tensor, the Ricci scalar  and the covariant derivative associated with the
{\it background} metric $\overline{g}_{\mu\nu}$.  The indices
are lowered or raised with  $\overline{g}_{\mu\nu}$  and $h \equiv \overline{g}^{\mu\nu} h_{\mu\nu} $ denotes the trace of the metric perturbation.
The linearised Einstein equations are then given by the Euler-Lagrange equations of (\ref{eq:perturbed-EH-action}) and can be written in the form
\bea
{\cal E}_{\mu\nu} &\equiv   &\bar\nabla_\sigma \bar\nabla^\sigma  h_{\mu\nu} + \bar\nabla_\mu \bar\nabla_\nu h + (\bar\nabla_\alpha \bar\nabla_\beta h^{\alpha\beta}- \bar\nabla_\sigma \bar\nabla^\sigma h) \overline{g}_{\mu\nu} + 2 \bar\nabla_{(\mu} \bar \nabla_\alpha h^\alpha_{\nu)}  -6 \bar\nabla_\alpha \bar\nabla_{(\mu} h_{\nu)}^\alpha   \nonumber \\
&+&  \bar R_{\mu\nu} h - \bar R h_{\mu\nu} + \frac{1}{2} \bar R \, \overline{g}_{\mu\nu} h + \bar R^{\alpha \beta}  \overline{g}_{\mu\nu}  h_{\alpha \beta} 
+ 8 \bar R_{\alpha(\mu} h^\alpha_{\nu)}=0 \, ,
\label{eom}
\eea
where use the standard notation $A_{(\mu\nu)} \equiv (A_{\mu\nu} + A_{\nu\mu})/2$ for the symmetrisation of any rank-2 tensor $A_{\mu\nu}$.

Let us now specialise these equations to the case where the background metric is the Schwarzschild metric,  expressed as 
\begin{equation}
\label{eq:schwarzschild-metric}
\tensor{\bar{g}}{_\mu_\nu} \dd\tensor{x}{^\mu} \dd\tensor{x}{^\nu} = -\left(1 - \frac{\mass}{r}\right) \dd{t}^2 + \left(1 - \frac{\mass}{r}\right)^{-1} \dd{r}^2 + r^2\left( \dd{\theta}^2 +  \sin^2 \theta  \dd{\varphi}^2\right)\,,
\end{equation}
where $\mass=2M_{s}$ is the Schwarzschild radius,   $M_s$ being the  mass of the black hole.

Given the spherical symmetry of the background solution, it is convenient to decompose the metric perturbations $h_{\mu\nu}$
into (scalar, vectorial and tensorial) spherical harmonics that are defined from the  standard $Y_{\ell m}(\theta, \varphi)$ functions
and their derivatives with respect to $\theta$ and $\varphi$.  They are labelled by the two multipole  integers $\ell$ and $m$ (with $\ell\geq 0$ and $-\ell \leq m \leq \ell$).

Furthermore, one can distinguish  axial and  polar modes, which behave differently under the parity transformation $\vec{r}\rightarrow -\vec{r}$: the polar, or even-parity, modes transform as $(-1)^\ell$, similarly to the scalar spherical harmonics $Y_{\ell m}(\theta, \varphi)$, whereas the axial, or odd-parity, modes transform as $(-1)^{\ell+1}$.  These modes can be  treated  separately as they are  decoupled at  linear order. Moreover, we consider here only the modes $\ell\geq 2$. The  particular cases of the  $\ell=0$ and $\ell=1$ modes  are briefly discussed in Appendix \ref{monopoleanddipole}. 

Since the background metric is static, it is also convenient to decompose the time dependence of the perturbations into Fourier modes, 
\begin{equation}
\label{eq:fourier}
F(t,r) = \int_{-\infty}^{+\infty} \dd\omega \, \tilde{F}(\omega, r)  e^{-i \omega t} \, .
\end{equation}
In the rest of this paper, we will use the same notation for the time-dependent function $F$ and its Fourier transform, as there will be no ambiguity. From a practical point of view, we simply  
 replace every time derivative by a multiplication by $-i \omega$ in the linearised equations, 
 which leads to a system of ordinary differential equations with respect to the radial variable $r$. 
 
 In  both  axial and polar sectors,  the equations of motion can be reduced to a system of two first order ordinary differential equations, as we will show below.

\subsubsection{Axial perturbations}
\label{Sec:OddPert}
We choose the usual Regge-Wheeler gauge \cite{Regge:1957td} to describe the axial modes. As recalled in  Appendix \ref{Odd-parity perturbations}, 
in this gauge the perturbations for $\ell\geq 2$ are parametrised by three families of functions $h_{0}^{\ell m}$, $h_{1}^{\ell m}$ and $h_{2}^{\ell m}$  according to
\bea
&&h_{t\theta} = \frac{1}{\sin\theta}  \sum_{\ell, m} h_0^{\ell m} \partial_{\varphi} {Y_{\ell m}}(\theta,\varphi), \qquad
h_{t\varphi} = - \sin\theta  \sum_{\ell, m} h_0^{\ell m} \partial_{\theta} {Y_{\ell m}}(\theta,\varphi), \nonumber \\
&&h_{r\theta} =  \frac{1}{\sin\theta}  \sum_{\ell, m} h_1^{\ell m}\partial_{\varphi}{Y_{\ell m}}(\theta,\varphi), \qquad
h_{r\varphi} = - \sin\theta \sum_{\ell, m} h_1^{\ell m}  \partial_\theta {Y_{\ell m}}(\theta,\varphi), \label{eq:odd-pert} 
\eea
while the other components  vanish.

For these perturbations,  the  equations of motion (\ref{eom}) reduce to the following three equations
\begin{equation}
\label{eq:full-eom-schwarzschild-odd}
\begin{aligned}
&{\cal E}_{t\theta} = 2 \left(\frac{\mass}{r} - 1 -\lambda \right) h_0(t,r) + r(r-\mass) \pdv[2]{h_0}{r} - 2(r-\mass) \pdv{h_1}{t} - r(r-\mass) \pdv[2]{h_1}{t}{r}=0  \, ,\\
&{\cal E}_{r\theta} = -2 \lambda \, h_1(t,r) - \frac{2r^2}{r-\mass} \pdv{h_0}{t} + \frac{r^3}{r-\mass} \pdv[2]{h_0}{t}{r} - \frac{r^3}{r-\mass} \pdv[2]{h_1}{t}=0 \,, \\
& {\cal E}_{\theta\theta} =2\mass h_1(t,r) + 2r(r-\mass) \pdv{h_1}{r} - \frac{2r^3}{r-\mass} \pdv{h_0}{t}=0  \, ,
\end{aligned}
\end{equation}
where 
we have introduced the notation
 \bea
2 \lambda \; \equiv \; \ell (\ell +1) - 2 \, ,
 \eea 
 as  the equations  ${\cal E}_{t\varphi}=0$, ${\cal E}_{r\varphi}=0$,  ${\cal E}_{\varphi\varphi}=0$  and ${\cal E}_{\theta\varphi}=0$ are identical to the above ones. 
 
Since there are  only two independent functions, $h_0$ and $h_1$, one expects one of the above equations to be redundant. This is indeed verified by noting  the following relation between the equations \eqref{eq:full-eom-schwarzschild-odd} and their derivatives, 
written now in the frequency domain,
\begin{equation}
\dv{\mathcal{E}_{r\theta}}{r} +\frac{i r^2 \omega}{(r-\mass)^2} \mathcal{E}_{t\theta} + \frac{\mass}{r(r-\mass)} \mathcal{E}_{r\theta} + \frac{\lambda}{r(r-\mass)} \mathcal{E}_{\theta\theta} = 0 \, .
\end{equation}
This shows that the two equations $\mathcal{E}_{r\theta}=0$ and $\mathcal{E}_{\theta\theta}=0$ are sufficient to fully  describe
the dynamics of axial perturbations. As a consequence, the initial system \eqref{eq:full-eom-schwarzschild-odd} reduces to
\bea
\label{eq:systeme-2-eqs-odd}
\dv{\X}{r} = M(r) \X \, , \quad
M(r) = 
\left(
\begin{array}{cc}
 {2}/{r} &  2i \lambda (r-\mass)/{r^3} - i \omega^2   \\
 -{i r^2}/{(r-\mass)^2} &  - {\mass}/{r(r-\mass)} 
\end{array}
\right) \, ,
\eea
where the two components of the column vector $\X \equiv {}^T\!(\X_1,\X_2)$ are $\X_1(r) \equiv h_0(r)$ and $\X_2(r)\equiv h_1(r)/\omega$.
Notice that we divided the variable $h_1(r)$ by $\omega$ in the definition of $\X_2$ in order to get a system which does not involve  powers of $\omega$ higher than 2, or equivalently which is at most  second order in time if one inverts the Fourier transform \eqref{eq:fourier}.

\subsubsection{Polar perturbations}
\label{Sec:EvenPert}
After fixing the gauge, polar perturbations  are parametrised by four families of functions $H_{0}^{\ell m}, H_{1}^{\ell m}, H_{2}^{\ell m}$ and $K^{\ell m}$ as shown in  Appendix \ref{Even-parity perturbations}. The nonvanishing metric perturbations then read
\bea
\label{eq:even-pert}
&&h_{tt} = A(r)\sum_{\ell, m} H_{0}^{\ell m}(t,r) Y_{\ell m}(\theta,\varphi), \quad
h_{tr} = \sum_{\ell, m} H_{1}^{\ell m}(t,r) Y_{\ell m}(\theta,\varphi),\\
&&h_{rr} = \frac{1}{A(r)} \sum_{\ell, m} H_{2}^{\ell m}(t,r) Y_{\ell m}(\theta,\varphi), \quad 
h_{ab} = \sum_{\ell, m} K^{\ell m}(t,r) g_{ab} Y_{\ell m}(\theta,\varphi),
\eea
where  $A(r) \equiv 1-\mass/r$ is included in the definitions for later convenience, and the indices $a$ or $b$ in the last equation 
are the angles $\theta$ or $\varphi$. 

The linearised Einstein's equations yield seven distinct equations, which can be found in  \eqref{eq:full-eom-schwarzschild} of Appendix \ref{App_Eq_even}. After a few 
manipulation, also discussed in Appendix \ref{App_Eq_even}, one finds that these  equations of motion 
can be reduced to two first-order equations only. In the frequency domain, they read
\bea
\label{eq:systeme-2-eqs}
\dv{\X}{r} = M(r) \X \,,\quad
M(r) = \frac{1}{3 \mass +2 \lambda  r}\begin{pmatrix}
	\frac{\mass  (3 \mass +(\lambda -2) r) - 2 r^4 \omega ^2}{r (r-\mass ) } & \frac{2 i (\lambda +1) (\mass
   +\lambda  r)+2 i r^3 \omega ^2}{r^2 } \\
 \frac{i r \left(9 \mass ^2-8 \lambda  r^2+8 (\lambda -1) \mass  r\right) + 4 i r^5 \omega ^2 }{2 (r-\mass )^2 } & \frac{2 r^4 \omega ^2-\mass  (3 \mass +3 \lambda  r+r)}{r (r-\mass )} \\
 \end{pmatrix} \
\eea
where now the two components of $\X$ are defined by $\X_1(r)\equiv K(r)$ and $\X_2(r) \equiv H_1(r) /\omega$. 
Similarly to the axial sector, 
the definition of $\X_2$ is motivated by the fact that  the resulting system involves at most $\omega^2$ terms.

\subsection{Schr\"odinger-like equation and effective potential}
\label{secondorderGR}

In both axial and polar sectors,  the equations of motion  have been recast in the form of a system consisting simply of two first-order differential  equations (with respect to the radial variable), namely \eqref{eq:systeme-2-eqs-odd} for axial perturbations and 
 \eqref{eq:systeme-2-eqs} for polar perturbations. In both cases, we now recall how this system can be rewritten as a Schr\"odinger-like equation.
 
 \subsubsection{From the first order system to the  Schr\"odinger-like equation}

 As shown in \cite{Regge:1957td} and \cite{Zerilli:1970se}, one can rewrite these systems as a single  second order (in radial derivatives)
Schr\"odinger-like equation for a unique dynamical variable. Reformulating a first order system of this kind 
as a Schr\"odinger equation is, in general, not an easy task because one has to ensure that the Schr\"odinger equation
is second order in time and in space. It requires, in particular, a decoupling of the dynamical variables involved in the original first order system and a ``clever''  choice for the dynamical variable that should satisfy the second order Schr\"odinger equation.

In this section, we will  describe how this works for the two systems  \eqref{eq:systeme-2-eqs-odd} and  \eqref{eq:systeme-2-eqs}  which
take the general form
\bea
\dv{\X}{r} = M(r) \X  \, ,
\label{eq:first-order-system}
\eea
where the coefficients of the matrix $M$ are polynomials (of degree at most 2) in $\omega$ and rational functions in $r$.

First, we consider the general (linear) change of vector
\bea
\label{eq:change-vars}
\X(r) = P(r) \hat \X (r) \, ,
\eea
where $\hat \X$ is a new column vector and the two dimensional invertible matrix $P$ has  not been fixed at this stage. We also define a new radial coordinate $r_*$ and introduce the ``Jacobian''
of the transformation $n(r) \equiv {{\rm d}r}/{{\rm d} r_*}$. Now, the idea is to show that it is possible to find a matrix $P$ such that the new system satisfied by 
$\hat{\X}$ takes the canonical form
\begin{equation}
\label{eq:systeme-reduit}
\dv{\hat \X}{r_*}  = \begin{pmatrix} 0 & 1 \\ V(r) - \omega^2 & 0\end{pmatrix} \hat \X \, ,
\end{equation}
where the potential $V(r)$ depends on  $r$, but not on $\omega$. Somehow, the first component $\hat{\X}_1$ plays the role of the ``momentum'' conjugate to the second component $\hat{\X}_2$ which would immediately implies that $\hat{\X}_1$ is the  ``canonical'' 
variable satisfying 
the required Schr\"odinger-like equation
\bea
\dv[2]{\hat{\X}_1}{r_*}  + \left(\omega^2 - V(r)\right) \hat{\X}_1 = 0 \, .
\label{eq:schrodinger-like-general}
\eea

\subsubsection{Axial modes}
Applying this procedure  to the system  \eqref{eq:systeme-2-eqs-odd} for the axial perturbations is rather simple\footnote{When one changes 
variables according to \eqref{eq:change-vars}, the new variable $\hat \X$ satisfies the differential equation
\bea
\dv{\hat \X}{r_*} = \hat M \hat \X \, , \qquad \hat M \equiv n(r) (P^{-1} M P - P^{-1} P') \, ,
\eea 
where $P'$ is the derivative of $P$ with respect to $r$, $M$  is the matrix introduced in \eqref{eq:systeme-2-eqs-odd} while  
$\hat M$ is the matrix entering in the system \eqref{eq:systeme-reduit}. 
They take a similar form $M= M_{[0]} + \omega^2 M_{[2]} $ and $\hat M= \hat M_{[0]}+  \omega^2 \hat M_{[2]} $ where
the expressions of $M_{[0]} $, $M_{[2]} $, $\hat M_{[0]} $ and $\hat M_{[2]} $ are trivially obtained.
As $P$ does not depend on $\omega$, the relation between $M$ and $\hat M$ translates into the two matricial relations
$\hat M_{[2]}  = n(r)P^{-1} M_{[2]}  P$ and $ \hat M_{[0]}  = n(r)(P^{-1} M_{[0]}  P - P^{-1} P' )\,$
which can be viewed as 8 equations for the 6 unknowns $n(r)$, $V(r)$ together with the four components of $P$.
Interestingly, the system is not overdetermined and admits a solution for $P$ \eqref{PGRodd}, for the potential $V(r)$ \eqref{VoddGR} and
for the function $n(r)$ which can be shown to be associated with the tortoise coordinate \eqref{eq:tortoise-coordinate}. Details can be found in the Appendix D of the companion paper.}. 
Indeed,  the appropriate transition matrix  is  given by
\bea
\label{PGRodd}
P(r) = \begin{pmatrix}
 1 - {\mass}/{r} & r \\
 -{i r^2}/{(r-\mass)} & 0 
\end{pmatrix}\,,
\eea
while $n(r) = 1 - {\mass}/{r}$, which means that $r_*$ coincides with the ``tortoise" coordinate,
\begin{equation}
r_* \equiv \int \frac{dr}{1 - {\mass}/{r}}= r + \mass \ln (r/\mass - 1) \, .
\label{eq:tortoise-coordinate}
\end{equation}
Finally the effective potential $V_\text{odd}(r) $ for the axial perturbations takes the form
\begin{equation}
\label{VoddGR}
V_\text{odd}(r) = \left(1 - \frac{\mass}{r}\right) \frac{2 (\lambda +1)r - 3\mass}{r^3}.
\end{equation}
Note that this potential vanishes both at spatial infinity ($r\rightarrow +\infty$) and at the horizon ($r\rightarrow \mass$).

\subsubsection{Polar modes}
The case of polar perturbations is slightly more involved. Starting from the system  \eqref{eq:systeme-2-eqs}, we find that the transition matrix  leading to a canonical form \eqref{eq:systeme-reduit} is given by\footnote{ We follow the same method as the one described in the previous footnote for the axial mode.}
\bea
\label{eq:change-vars-gr-even}
P = \begin{pmatrix}
 \frac{3 \mass^2 + 3 \lambda \mass r  + 2 r^2 \lambda (\lambda +1)}{2r^2 (3\mass + 2 \lambda r)} & 1 \\
 -i +  \frac{i \mass}{2(r-\mass)} + \frac{3 i \mass}{3\mass + 2 \lambda r }  & -\frac{i r^2}{r-\mass} 
\end{pmatrix} \, ,
\eea 
with, in addition, $n(r) = 1 - {\mass}/{r}$, which means that $r_*$ is still the tortoise coordinate \eqref{eq:tortoise-coordinate}. 
Finally,  the corresponding potential $V_\text{even}(r)$ reads
\begin{equation}
\label{eq:potential-schwarzschild}
V_\text{even}(r) = \left(1-\frac{\mass}{r}\right) \frac{9\mass^3 + 18 \mass^2 r\lambda + 12\mass r^2 \lambda^2 +8  r^3 \lambda^2(1+\lambda)}{r^3 (3\mass + 2r\lambda)^2} \, .
\end{equation}

\begin{figure}[h]
 \captionsetup{singlelinecheck = false, format= hang, justification=raggedright, font=footnotesize, labelsep=space}
	\includegraphics{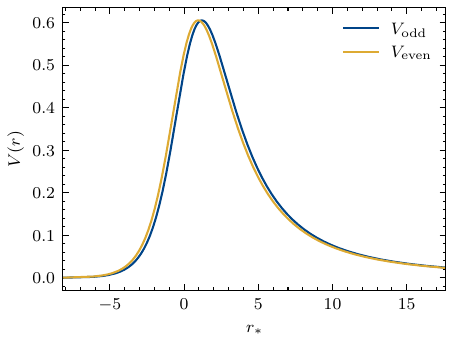}
	\caption[left]{Illustration of the effective potentials (for axial and polar modes) for a Schwarzschild black hole. The parameters are such that 
	$\mass=2$ (i.e. the mass of the black is $1$ in natural units) and $\ell=2$ here.}
	\label{GRpotentials}
\end{figure}

Despite their different analytic forms, we notice in Fig.\eqref{GRpotentials} that the potentials $V_\text{odd}(r)$ and $V_\text{even}(r)$ are quite similar, although distinct. In fact, there exists an underlying symmetry between these two potentials, further explained in \cite{Chandrasekhar:1985kt}, leading to the isospectrality theorem which states that the spectra
of axial and polar perturbations are  exactly the same.  

\subsubsection{Quasi-normal modes and boundary conditions}

Finding quasi-normal modes requires to impose the appropriate boundary conditions: the modes must be outgoing at infinity and ingoing at the horizon. 

Since  both $V_\text{odd}$ and $V_\text{even}$ go to zero at infinity and at the horizon,  equation \eqref{eq:schrodinger-like-general} becomes asymptotically
\begin{equation}
	\dv[2]{\hat{X}_1}{r_*} + \omega^2 \hat{X}_1 \approx 0 \qquad (r_*\rightarrow \pm\infty)\,,
\end{equation}
where $ \approx$ is an equality up to sub-leading corrections\footnote{Near the horizon, $V = {\cal O}(r-\mass)$ for both potentials, hence we assume $\mass^2 \omega^2 \gg r/\mass-1$. At infinity,  $V = {\cal O}(1/r^2)$ for both potentials as well, hence we assume $\omega^2 r^2 \gg 1$ in this limit.}.
Therefore, at both boundaries, the function $\hat{X}_1$ behaves like 
\bea
\hat{X}_1(r) \approx  {\cal A} \, e^{i\omega r_*} + {\cal B} \, e^{-i\omega r_*}  \,,
\eea
where  ${\cal A}$ and ${\cal B} $ are integration constants which take different values at the horizon and at infinity.

 The physical interpretation of these modes is more transparent if we include their time dependence explicitly, which gives 
\begin{equation}
	\left\{
	\begin{aligned}
		\hat{X}_1(t,r) & \approx {\cal A}_\text{hor}\,  e^{-i \omega (t - r_*)} + {\cal B} _\text{hor}\, e^{-i \omega (t + r_*)}  \quad \text{when} \,\, {r \longrightarrow \mass}, \\
		\hat{X}_1(t,r) & \approx {\cal A}_\infty \, e^{-i \omega (t - r_*)} + {\cal B} _\infty\,  e^{-i \omega (t + r_*)}  \quad \text{when} \,\, {r \longrightarrow \infty} \,.
	\end{aligned}
	\right.
\end{equation}
We can interpret each term as a radially propagating wave: the terms proportional to ${\cal A}_\text{hor}$ and ${\cal A}_\infty$ are outgoing while the terms proportional to ${\cal B} _\text{hor}$ and ${\cal B} _\infty$ are ingoing. Imposing a  purely outgoing behaviour at infinity and a purely ingoing behaviour at the horizon, i.e.  such that ${\cal A}_\text{hor} = 0$ and ${\cal B} _\infty = 0$ severely restricts  the possible values of $\omega$. These values can be found numerically by integrating the Schr\"odinger-like equation (see   \cite{Leaver:1985ax} and the reviews \cite{Kokkotas:1999bd,Nollert:1999ji,Berti:2009kk,Konoplya:2011qq}).

Finally, one can easily deduce the asymptotic expansion of the original gravitational perturbations using the transformations \eqref{eq:change-vars}. For the axial modes, the leading order terms at infinity are thus given by
\bea
h_0(r)  \approx  i \omega r \left( {\cal A}_\infty e^{i \omega r_*} - {\cal B} _\infty e^{-i\omega r_*}\right)  , \quad 
h_1(r)   \approx  - i \omega r \left( {\cal A}_\infty e^{i \omega r_*} + {\cal B} _\infty e^{-i\omega r_*}\right)    \label{asymptVoddH1inf} \, ,
\eea
while the leading order terms at the horizon read
\bea
h_0(r)   \approx  i \omega \mass  \left( {\cal A}_\text{hor} e^{i \omega r_*} - {\cal B} _\text{hor}  e^{-i\omega r_*}\right)   \,  , \quad
h_1(r)   \approx  - \frac{i \omega \mass^2}{\varepsilon}   \left( {\cal A}_\text{hor}  e^{i \omega r_*} + {\cal B} _\text{hor} e^{-i\omega r_*}\right)  \, ,\label{asymptVoddH1hor} 
\eea
where we have introduced the variable $\varepsilon \equiv r - \mass$ which satisfies $\varepsilon \ll \mass$ near the horizon. 

For the polar modes, the leading order terms at infinity are
\bea
K(r)  \approx i \omega  \left( {\cal A}_\infty e^{i \omega r_*} - {\cal B} _\infty e^{-i\omega r_*}\right)  \, , \quad
H_1(r) \approx   r \omega^2 \left( {\cal A}_\infty e^{i \omega r_*} - {\cal B} _\infty e^{-i\omega r_*}\right)   \label{asymptVevenH1inf} \, ,
\eea
while the leading terms at the horizon are a bit more involved and read 
\bea
K(r) & \approx &  \frac{\lambda +1 + 2 i \omega \mass}{\mass}  {\cal A}_\text{hor}  e^{i \omega r_*} + \frac{\lambda +1 - 2 i \omega \mass}{\mass}  {\cal B} _\text{hor}  e^{-i \omega r_*}  \label{asymptVevenKhor} \, , \\
H_1(r) & \approx & \frac{i \mass \omega(1-2i\omega \mass)}{2\varepsilon}  {\cal A}_\text{hor}  e^{i \omega r_*}  + 
 \frac{i \mass \omega(1+2i\omega \mass)}{2\varepsilon} {\cal B} _\text{hor}  e^{-i \omega r_*}   \, .  \label{asymptVevenH1hor}
\eea
In the next section, we will recover these asymptotic behaviours in a completely different way.

\section{First order approach to Schwarzschild perturbations}
\label{sec:Schwarzschild_first_order}

As we have seen in  \autoref{secondorderGR}, finding a (second-order) Schr\"odinger-like equation for the metric perturbations starting 
from the Einstein equations requires some manipulations of the equations of motion and an appropriate choice of the function that verifies the Schr\"odinger-like equation.

The rest of this paper will be devoted to obtaining the asymptotic behaviours of the perturbations by using a different method. Although this is of course not necessary for the perturbations of Schwarzschild in General Relativity, our method may prove to be very useful in situations where a Schr\"odinger-like system is not obvious to find or  even impossible to reach. In such a case, one would need an alternative method to determine the asymptotic limits of the solutions of the system, and from them, to compute the quasi-normal modes.

The general method  will be described in a systematic  way in the next section. As the general procedure is somewhat tedious, we have preferred to present it first, in a pedestrian way, for the perturbations of Schwarzschild. A more mathematically-minded reader might prefer to jump directly to the next section and later come back to this section to find a particular application of the general method.

\subsection{Method}
\label{sec:Schwarzschild_first_order_method}
Ignoring the traditional Schr\"odinger reformulation,  we now go back to the original first-order system given in  \eqref{eq:systeme-2-eqs-odd} or \eqref{eq:systeme-2-eqs}.
Schematically,  we thus have a  first-order system of the form 
\bea
\label{system}
 \dv{{\X}}{r} =  M(r)  \X \, , \qquad 
\eea
where $\X(r)$ is a column vector and  $M(r)$ a square matrix.
In order  to study the system  at  spatial infinity, say, i.e. when $r\rightarrow \infty$, one can expand the matrix $M(r)$ in powers of $r$,
\beq
\label{M_system}
 M(r) = M_p\, r^p +\dots + M_0+M_{-1}\, \frac1r+ {\cal O}(\frac{1}{r^2})
\eeq
where all the matrix coefficients $M_i$ are $r$-independent. We stop here the expansion at order $1/r$, which is sufficient for the simplest cases, but  higher orders might be needed in general.

If {\it all} matrices $M_i$ are {\it diagonal}, it is  immediate to integrate the truncated system, which then consists of $n$ ordinary differential equations of the form
\beq
y'(r)=\left(\lambda_p r^p+\dots+\lambda_0 +\frac{\mu}{r}\right)y(r)\,,
\eeq
whose  solution is 
\beq
 y(r)= y_0\,  e^{q(r)} r^{\mu}\,, \quad q(r)= \frac{\lambda_p}{p+1} r^{p+1}+\dots+\lambda_0 r\,.
\eeq
Putting together these $n$ solutions, we thus get the solution to the system (\ref{system}), assuming all matrices $M_i$ in \eqref{M_system} are diagonal, in the form
\beq
\label{asymptotic}
\X(r)= e^{\mathbf{\Upsilon}(r)} r^\mathbf{\Delta} \mathbf{F}(r) \X_0
\eeq
where $\X_0$ is a constant vector, corresponding to the $n$ integration constants, $\mathbf{\Upsilon}$ is a diagonal matrix whose coefficients are polynomials of degree at most $p+1$, $\mathbf{\Delta}$ is a constant diagonal matrix and $\mathbf{F}(r)$ is a matrix which is regular at infinity  (i.e. whose limit is finite).

 Of course, in general, the matrices $M_i$ are not diagonal but, remarkably, it is always possible to transform the truncated system into a fully diagonal system, in a {\it finite} number of steps following an algorithm introduced in \cite{wasow_asymptotic_1965,balser_computation_1999,barkatou_algorithm_1999,pflugel_root-free_2019}, which we will present in full details in the next section.

At  each step in the algorithm, one introduces a new vector $\tilde \X$, related to the vector  $\X$  of the previous step by
\begin{equation*}
	\X = P \tilde{\X} \,,
\end{equation*}
where $P$ is an invertible matrix
so that the  previous system \eqref{system}  is transformed into a new, but equivalent, system of the form
\begin{align}
	\dv{\tilde{\X}}{r} = \tilde{M}(r) \tilde{\X}, \qquad
	\tilde{M}(r) \equiv P^{-1} M P - P^{-1} \dv{P}{r} \,.
	\label{eq:transfo-M_0}
\end{align}
The idea is then to choose an  appropriate transition matrix $P$ at each step in order to diagonalise, order by order, the matrices that appear in the expansion of $M$. Once all the matrices are diagonalised, one can integrate directly the diagonal system, as we have seen earlier,  and obtain the general asymptotic solution of the system.

For the asymptotic behaviour near the horizon, one proceeds in the same way by noting that the variable $z=1/(r-\mass)$ goes to infinity when $r\rightarrow \mass$. In the rest of this section, we will illustrate the algorithm by considering in turn the asymptotic behaviours of the axial and polar modes.

\subsection{Axial modes}
\label{subsection_axial}
The analysis of the asymptotic behaviour of the first order system \eqref{eq:systeme-2-eqs-odd} is relatively simple and  instructive. We recall that the system is of the form 
\bea
\label{generalsystem}
 \dv{{\X}}{r} = {M}(r) {\X} \, , 
\eea
with
\bea
\X(r) \equiv
\left(
\begin{array}{c}
h_0(r) \\ h_1(r)/\omega
\end{array}
\right) \, , \qquad
M(r) \equiv
\left(
\begin{array}{cc}
 {2}/{r} &  2i \lambda (r-\mass)/{r^3} - i \omega^2   \\
 -{i r^2}/{(r-\mass)^2} &  - {\mass}/{r(r-\mass)} 
\end{array}
\right) \, .
\label{oddMmatrix}
\eea

\subsubsection{Asymptotic analysis at spatial infinity}
We first study the asymptotic behaviour at spatial infinity, i.e. when $r \rightarrow \infty$. 
 The asymptotic expansion of the matrix $M(r)$ at large $r$ reads
 \bea
M(r) = M_0 + \frac{1}{r} M_{-1} + {\cal O}\left( \frac{1}{r^2}\right) \, , \qquad
M_0 \equiv -i\left(
\begin{array}{cc}
0 & \omega^2 \\
1 & 0
\end{array}
\right) \, , \quad 
M_{-1} \equiv {2} 
\left(
\begin{array}{cc}
1 & 0 \\
-i \mass  & 0
\end{array}
\right)  \, .
\eea
The leading term $M_0$  is diagonalisable and one can go to a basis where it is diagonal, by introducing the new vector $\X^{(1)}$ defined
 by
\bea
\X \equiv P_{(1)}  {\X}^{(1)}\,,\qquad 
P_{(1)} = \left( \begin{array}{cc} \omega & -\omega \\ 1 & 1 \end{array}\right)  \,.
\eea
According to \eqref{eq:transfo-M_0},  this gives the new system
\bea
 \dv{{\X}^{(1)}}{r} = {M}^{(1)} {\X}^{(1)} \, , \quad 
	{M}^{(1)}(r) = {M}^{(1)}_0 + \frac{1}{r} {M}^{(1)}_{-1} + {\cal O}\left( \frac{1}{r^2}\right) \, , 
\eea
with
\bea	
  {M}^{(1)}_0 \equiv  \left( \begin{array}{cc} -i \omega & 0 \\ 0 &i \omega \end{array}\right) \, , \quad
	 {M}^{(1)}_{-1} \equiv \left( \begin{array}{cc} -i   \omega\mass  +1 & i \omega \mass -1\\ -i \omega \mass - 1 & i \omega\mass  + 1 \end{array}\right) . \label{OddEq1}
\eea
We need some extra work to diagonalise the next-to-leading order matrix ${M}^{(1)}_{-1}$ while keeping the leading order matrix diagonal. 

This can be achieved by introducing a new  vector $\X^{(2)}$ defined by 
\bea
\X^{(1)} \equiv P_{(2)} \X^{(2)}\,, \qquad 
P_{(2)} \; =  \Id  + \frac{1}{r} \Xi \, ,
\eea
where  $\Id$ is the identity matrix and $\Xi$ a constant matrix.
Indeed, it is immediate to see that such a change of variable leads to the  equivalent  differential system,
\bea
  \dv{{\X}^{(2)}}{r} = {M}^{(2)} {\X}^{(2)} \, , \quad 
	{M}^{(2)}(r) = {M}^{(2)}_0 + \frac{1}{r} {M}^{(2)}_{-1} + {\cal O}\left( \frac{1}{r^2}\right) \, , 
\eea
with
\bea	
	 {M}^{(2)}_0  = {M}^{(1)}_0 \, , \quad  {M}^{(2)}_{-1} =  {M}^{(1)}_{-1}  + [ {M}^{(1)}_{0},\Xi ] \,. \label{OddEq2}
\eea
The leading matrix remains unchanged while one can easily find a matrix $\Xi$ so that ${M}^{(2)}_{-1}$ is diagonal. Notice that 
$\Xi$ appears in \eqref{OddEq2} only in a commutator with the diagonal matrix $M_0^{(1)}$, hence the diagonal part of $\Xi$ is
irrelevant and we can already fix the diagonal terms of $\Xi$  to $0$. In this case, the solution to  \eqref{OddEq2}  with ${M}^{(2)}_{-1}$  diagonal is unique and given by
\bea
\Xi \; = \; \frac{1}{2i\omega}
 \left (
\begin{array}{cc}
 0 &  i  \omega\mass -1\\
   { i  \omega \mass + 1} & 0
\end{array}
\right) \, .
\eea

We have thus managed to obtain a fully diagonalised system, up to order $1/r$, with the matrix 
\bea
{M}^{(2)}(r) =  \left( \begin{array}{cc} -i \omega & 0 \\ 0 &i \omega \end{array}\right) + 
	\frac{1}{ r}\left( \begin{array}{cc} 1-i \omega \mass & 0\\ 0 & 1+i \omega\mass   \end{array}\right) + {\cal O}\left( \frac{1}{r^2}\right).
\eea
This system can be immediately integrated in the form \eqref{asymptotic}, and the asymptotic solution reads
\bea
{\X}^{(2)}(r)  \; = \; \left(1 + {\cal O}\left( {1}/{r}\right)\right) \left(
\begin{array}{c}
c_-   \, e^{-i\omega r} r^{1- i \omega\mass  } \\
c_+ \, e^{+i\omega r} r^{1+ i  \omega\mass   }
\end{array}
\right) \, ,
\label{AsymptoticOdd}
\eea
where $c_\pm$ are integration constants. Taking into account the time dependency $e^{-i\omega t}$ of the modes, the two components $\X^{(2)}_\mp$ of $\X^{(2)}$ are
of the form
\bea
e^{-i \omega t} \, \X^{(2)}_\mp(r) = \left(1 + {\cal O}\left( {1}/{r}\right)\right) c_\mp r  e^{-i \omega (t \pm ( r +\mass \ln r))}= c_\mp \left(r + {\cal O}(1)\right) e^{-i \omega (t \pm r_*)} \, ,
\eea
where it is convenient to use the ``tortoise'' coordinate $r_*$, introduced in \eqref{eq:tortoise-coordinate}, noting that
\bea
\label{tortoise}
r_* = r + \mass \ln (r/\mass - 1) = r + \mass \ln r + {\cal O}(1) \, .
\eea 
As a consequence, one can identify $\X^{(2)}_-$ as an ingoing mode and $\X^{(2)}_+$ as an outgoing mode at spatial infinity.

Finally, we can return to the original vector $\X$ thanks to the transformation
\bea
\X = P_{(1)} P_{(2)} \X^{(2)} = \left( \begin{array}{cc} \omega & - \omega \\ 1 & 1\end{array} \right) \left( 1 + \frac{\Xi}{r} \right) \X^{(2)} \, ,
\eea
in order to obtain the asymptotic expansion of the two original gravitational perturbations $h_0$ and $h_1$ at spatial infinity,
\bea
h_0(r) & = & \omega \left( c_- e^{- i \omega r_*} - c_+ e^{+ i \omega r_*}  \right)  \left(r + {\cal O}(1)\right) \, , \\
h_1(r) & = & \omega \left( c_- e^{- i \omega r_*} + c_+ e^{+ i \omega r_*}  \right)  \left(r + {\cal O}(1)\right) \, .
\eea
One can immediately check that  these expressions agree with the asymptotic expansion 
 \eqref{asymptVoddH1inf} obtained from the Schr\"odinger-like equation (with  $c_- = -i {\cal B}_\infty$ and $c_+ = -i {\cal A}_\infty$).

\subsubsection{Asymptotic analysis near the black hole horizon}

Let us now study the behaviour of the axial modes near the horizon. In this case, it is convenient to introduce the new radial variable $\varepsilon \equiv r-\mass$ and expand the matrix $M$ for the system \eqref{oddMmatrix}  in powers of $\varepsilon$. One finds\footnote{Note that $\varepsilon$ goes to zero here, in contrast to the previous case where the variable $r$ was going to infinity. One could work in a fully analogous system by using the variable $z=1/\varepsilon$, with the system
\beq
\frac{d\X}{dz}=\tilde M(z) \X\,, \quad \tilde M=-\frac{1}{z^2}M(z^{-1})=-M_2- M_1\frac{1}{z}-M_0\frac{1}{z^2}\,.
\eeq
In the present case, one must push the expansion up to order $1/z^2$ because the leading matrix $M_2$ is nilpotent.
}
\bea
M(\varepsilon) = \frac{1}{\varepsilon^2}M_{2} + \frac{1}{\varepsilon} M_1  + M_0 + {\cal O}(\varepsilon) \, , 
\eea
with the matrix coefficients
\bea
M_2 \equiv \left(  \begin{array}{cc}0 & 0 \\ -i\mass^2& 0\end{array} \right)\, , \quad
M_1 \equiv \left( \begin{array}{cc} 0 & 0 \\ -2 i \mass & -1 \end{array} \right)\, , \quad
M_0 \equiv \left(  \begin{array}{cc}2/\mass & -i \omega^2\\ -i & 1/\mass \end{array} \right) \, .
\eea
An important difference with the previous situation is that the leading term $M_2$ is no longer diagonalisable but  nilpotent instead.  We thus need to first perform a transformation that yields a diagonalisable leading matrix, taking advantage of the derivative term in \eqref{eq:transfo-M_0}. This can be done with the transformation 
\bea
\label{changeatHodd}
\X \equiv  P_{(1)} \X^{(1)}  \, , \quad P_{(1)}(\varepsilon) \equiv \left( \begin{array}{cc} 1 & 0 \\ 0 & 1/\varepsilon \end{array}\right) \,,
\eea
leading to the new system
\bea
\label{system_axial_horizon}
 \dv{{\X}^{(1)}}{\varepsilon} = {M}^{(1)} {\X}^{(1)} \, , \quad 
	{M}^{(1)}(\varepsilon) = -\frac{1}{\varepsilon} \left( \begin{array}{cc} 0 & i \omega^2 \\  i \mass^2 & 0\end{array}\right) + {\cal O}(1)\, .
\eea
The transformation \eqref{changeatHodd} has eliminated the term in $1/\varepsilon^2$ in the expansion and the leading term $M^{(1)}_1$ is now diagonalisable, so that only the expansion  of $M^{(1)}$ up to order $1/\varepsilon$ is required (see discussion in the footnote).
 It is worth noticing that $M^{(1)}_1$ receives contributions from
$M_2$, $M_1$ and $M_0$. In particular, some of its coefficients involve the frequency $\omega$ which is originally present only in $M_0$.

The final step of the analysis  consists in diagonalising the system \eqref{system_axial_horizon}, via the transformation
\bea
\label{changeatHodd2}
\X^{(1)} = P_{(2)} \X^{(2)} \, , \quad P_{(2)} \equiv \left( \begin{array}{cc} \omega & -\omega \\ \mass & \mass \end{array}\right) \, ,
\eea
leading to 
\bea
 \dv{{\X}^{(2)}}{\varepsilon} =M^{(2)} \X^{(2)} \, , \qquad  M^{(2)}(\varepsilon)  \equiv 
 \frac{1}{\varepsilon} \left( \begin{array}{cc} -i\omega\mass & 0\\ 0 & i\omega\mass\end{array}\right) + {\cal O}(1) \, .
\eea
Integrating this equation yields
\bea
\label{asympOddH}
{\X}^{(2)} (\varepsilon) = (1+ {\cal O}(\varepsilon)) \left( \begin{array}{c} c_- \varepsilon^{-i \omega\mass } \\ c_+ \varepsilon^{+i  \omega\mass } \end{array}\right) 
=  (1+ {\cal O}(\varepsilon))  \left( \begin{array}{c} c_- e^{-i \omega r_*}\\ c_+ e^{+i \omega r_*} \end{array}\right) \, ,
\eea
where we have again expressed the result in terms of   the tortoise coordinate $r_*$, which behaves as $r_* = \mass \ln \varepsilon +{\cal O}(1)$ near the horizon. One can immediately recognize the ingoing and outgoing modes at the horizon.

Finally, one can return to the original  functions, via $\X=P_{(1)}P_{(2)}\X^{(2)}$, and  derive the expressions 
\bea
h_0(r) & = &  \omega \left( c_- e^{- i \omega r_*} - c_+ e^{+ i \omega r_*}  \right)  \left(1 + {\cal O}(\varepsilon)\right) \, , \\
h_1(r) & = & \frac{\omega \mass }{\varepsilon} \left( c_- e^{- i \omega r_*} + c_+ e^{+ i \omega r_*}  \right)  \left(1 + {\cal O}(\varepsilon)\right) \, ,
\eea
which coincide with the asymptotic expansions
 \eqref{asymptVoddH1hor} obtained from the Schr\"odinger-like equation 
(with $c_-=-i \mass {\cal B}_\text{hor}$, $c_+=-i \mass {\cal A}_\text{hor}$).

\subsection{Polar modes}
The dynamics of the polar perturbations is  described by the first-order system \eqref{eq:systeme-2-eqs}, of  the form
\bea
\label{systemevengen}
 \dv{{\X}}{r} = {M}(r) {\X} \, ,  \qquad \text{with} \qquad \X(r) \equiv
\left(
\begin{array}{c}
K(r) \\ H_1(r)/\omega
\end{array}
\right) \, , 
\eea
and  the matrix
\bea
M(r) = \frac{1}{3 \mass +2 \lambda  r}\begin{pmatrix}
	\frac{\mass  (3 \mass +(\lambda -2) r) - 2 r^4 \omega ^2}{r (r-\mass ) } & \frac{2 i (\lambda +1) (\mass
   +\lambda  r)+2 i r^3 \omega ^2}{r^2 } \\
 \frac{i r \left(9 \mass ^2-8 \lambda  r^2+8 (\lambda -1) \mass  r\right) + 4 i r^5 \omega ^2 }{2 (r-\mass )^2 } & \frac{2 r^4 \omega ^2-\mass  (3 \mass +3 \lambda  r+r)}{r (r-\mass )} \\
 \end{pmatrix} \, . \label{Meven}
\eea
\subsubsection{Asymptotic analysis at spatial infinity}
Expanding \eqref{Meven} in powers of $r$, one gets 
\begin{eqnarray}
	M(r) & = & \begin{pmatrix} 0 & 0 \\ \frac{i \omega^2}{\lambda} & 0 \end{pmatrix} r^2
	+ \begin{pmatrix} - \frac{\omega^2}{\lambda} & 0 \\ \frac{i \mass\omega^2 (4\lambda - 3)}{2 \lambda^2} & \frac{\omega^2}{\lambda} \end{pmatrix} r
	+ \begin{pmatrix} -\frac{(2 \lambda -3) \mass  \omega ^2}{2 \lambda ^2} & \frac{i \omega ^2}{\lambda } \\
 -2 i+\frac{3 i \left(4 \lambda ^2-4 \lambda +3\right) \mass ^2 \omega ^2}{4 \lambda ^3} & \frac{(2 \lambda -3) \mass  \omega ^2}{2 \lambda ^2} \\ \end{pmatrix} \nonumber \\ 
&+&  \frac{1}{r}\begin{pmatrix} -\frac{\left(4 \lambda ^2-6 \lambda +9\right) \mass ^2 \omega ^2}{4 \lambda ^3} & -\frac{3 i \mass  \omega ^2}{2 \lambda ^2} \\  \frac{i \left(8 (1-2 \lambda ) \lambda ^3 \mass -(27-4 \lambda  (\lambda  (8 \lambda -9)+9)) \mass ^3 \omega ^2\right)}{8 \lambda ^4} & \frac{\left(4 \lambda ^2-6
		\lambda +9\right) \mass ^2 \omega ^2}{4 \lambda ^3} \end{pmatrix} 
+ \mathcal{O}\left(\frac{1}{r^2}\right) \, .
\end{eqnarray}
In contrast with the axial modes at spatial infinity, the leading matrix is of order $r^2$ and is nilpotent. So, in principle, one needs to apply a procedure similar to the near-horizon analysis of axial modes, which will be presented in full generality in the next section, and then diagonalise in turn all subsequent orders. All this involves many steps which are straightforward but rather tedious to describe. 

 To shorten our discussion, we provide directly the transformation that combines all these intermediate steps, given by 
\begin{equation}
\label{Peventotal}
\X=P\tilde{\X}\,, \qquad 
    P = \begin{pmatrix}
    {\cal S}  + {\cal T} & {\cal S}  - {\cal T}  \\
   {\cal U} - {\cal V}  &{\cal U} + {\cal V} 
\end{pmatrix},
\end{equation}
with the functions 
\begin{eqnarray}
                &&{\cal S}(r) \equiv \frac{i (r-\mass ) ((2 \lambda -3) \mass +4 \lambda  r)}{4 \lambda  r} +\frac{i \lambda }{2 r \omega^2}\, , \quad
            {\cal T}(r) \equiv \frac{(1-2 \lambda)  \mass  +2(1+2 \lambda)  r }{4 r \omega} \, , \nonumber \\
            &&
            {\cal U}(r) \equiv r^2+ \frac{2\lambda - 3}{4\lambda } \mass  r \, , \quad
            {\cal V}(r) \equiv \frac{ir}{2\omega} \, .
\end{eqnarray}%
This leads to the new system
\begin{equation}
 \dv{{\tilde{\X}}}{r} = \tilde{M}(r)  \tilde{\X} \, , \qquad
	\tilde{M}(r) = \begin{pmatrix}i \omega  & 0 \\
 0 & -i \omega  \\\end{pmatrix} + \begin{pmatrix}
 -1+i  \omega \mass   & 0 \\
 0 & -1-i   \omega\mass   \\
\end{pmatrix} \frac{1}{r} + \mathcal{O}\left(\frac{1}{r^2}\right),
\end{equation}
which is diagonal and whose solution is 
\begin{equation}
\tilde{\X}(r)  \; = \;
\left(
\begin{array}{c}
c_- \,  e^{-i\omega r} r^{-1- i \omega\mass  } \\
c_+ \, e^{+i\omega r} r^{-1+ i  \omega \mass }
\end{array}
\right)   \left(1 +  {\cal O} \left( {1}/{r}\right)\right)=
 \frac{1}{r}
\left(
\begin{array}{c}
c_- \,  e^{-i\omega r_*}  \\
c_+ \, e^{+i\omega r_*}   
\end{array}
\right)\left(1 + {\cal O}\left( {1}/{r}\right)\right)
\, .
\end{equation}
This result is very similar to that obtained for axial perturbations \eqref{AsymptoticOdd}, even though the asymptotic expansion of the matrix $M$ is rather different. 
In terms of  the original functions,   we find 
\bea
K(r)= \frac{i}{\omega} H_1(r)= i (c_-  \,  e^{-i\omega r_*} + c_+  \,  e^{+ i\omega r_*} ) \left(1 +  {\cal O} \left( {1}/{r}\right)\right) \,,
\eea 
which agree with \eqref{asymptVevenH1inf} (with $c_- = - \omega {\cal B}_\infty$ and $c_+= \omega {\cal A}_\infty$).

\subsubsection{Asymptotic analysis at the black hole horizon}

We finally turn to the near-horizon behaviour of polar modes. The expansion of the matrix \eqref{Meven} in terms of  the small parameter $\varepsilon\equiv r-\mass$
yields
\bea
M(\varepsilon)=\frac{1}{\varepsilon^2} M_2 + \frac{1}{\varepsilon} M_1 + M_0  + {\cal O}(\varepsilon) , \quad
M_2 = \begin{pmatrix} 0 & 0 \\ \gamma_2 & 0 \end{pmatrix}  , \;  
M_1 = \begin{pmatrix} \alpha_1 & 0 \\ \gamma_1 & \delta_1 \end{pmatrix} , \;  
M_0 = \begin{pmatrix}  \alpha_0 & \beta_0 \\ \gamma_0 & \delta_0 \end{pmatrix} , 
\eea
where only a few of the coefficients $ \alpha_I$, $\beta_I$ and $\gamma_I$ will be  needed explicitly. 

Once more, the dominant
$M_2$ is a nilpotent matrix and, as in  the axial case,
we use the transformation
\bea
\label{changeatEven}
\X=P_{(1)}\X^{(1)}   \quad \text{with} \quad P_{(1)}(\varepsilon) \equiv \left( \begin{array}{cc} 1 & 0 \\ 0 & 1/\varepsilon \end{array}\right) \, ,
\eea
which gives the new system
\bea
 \dv{{\X}^{(1)}}{\varepsilon} = {M}^{(1)} {\X}^{(1)} \, , \quad 
	{M}^{(1)}(\varepsilon) = \frac{1}{\varepsilon} \left( \begin{array}{cc}  \alpha_1 & \beta_0 \\ \gamma_2 & 1+ \delta_1 \end{array}\right) + {\cal O}(1)\, ,
\eea
with the coefficients
\begin{eqnarray}
 \alpha_1= -(1+\delta_1) = \frac{1+\lambda - 2 \mass^2 \omega^2}{3 + 2 \lambda} \, , \quad
\beta_0=\frac{2i}{\mass^2} \frac{(\lambda+1)^2+ \mass^2 \omega^2}{3 + 2 \lambda} \, , \quad
\gamma_2 = \frac{i \mass^2}{2} \frac{1+ 4\mass^2 \omega^2}{3 + 2 \lambda} \, .
\end{eqnarray}
The leading matrix can now be diagonalised via the transformation
\bea
\X^{(1)} = P_{(2)} \X^{(2)} \,, \quad \text{with} \quad P_{(2)} = \begin{pmatrix}  \alpha - \beta &  \alpha + \beta \\ 1 & 1\end{pmatrix} \quad\text{and}\quad  \alpha = \frac{ \alpha_1}{\gamma_2} \,, \quad \beta = \frac{i\omega\mass}{\gamma_2} \,.
\eea
leading to the system
\begin{equation}
	\dv{\X^{(2)}}{\varepsilon} = M^{(2)} \X^{(2)} \,,\quad M^{(2)} = \frac1\varepsilon \begin{pmatrix} - i  \omega \mass& 0 \\ 0 & i \omega\mass \end{pmatrix} + \mathcal{O}(1) \,.
\end{equation}
Note that this expression is extremely simple and does not involve $\lambda$, as expected, even though it appears explicitly in $M^{(1)}$. We obtain immediately the asymptotic behaviour of $X^{(2)}$ near the horizon
\begin{equation}
	\X^{(2)}(\varepsilon) = (1 + \mathcal{O}(\varepsilon)) \begin{pmatrix} c_- \, e^{-i \omega r_*} \\ c_+ \, e^{+i \omega r_*} \end{pmatrix} \,,
\end{equation}
which reproduces the same result as for the axial mode \eqref{asympOddH}.
In terms  of the original gravitational functions $H_1(r)$ and $K(r)$, using  the transformation $\X=P_{(1)} P_{(2)} \X^{(2)}$, we recover the result  \eqref{asymptVevenH1hor}, with
\beq
	c_+=\frac{i}{2}\mass (1 - 2i\omega \mass) {\cal A}_\text{hor}\qquad
	c_-=-\frac{i}{2}{ \mass (1 + 2i\omega \mass)} {\cal B}_\text{hor} \,.
\eeq
This completes our study of all asymptotic behaviours of Schwarzschild perturbations, demonstrating that one can recover the standard results directly from the linearised Einstein's equations, without resorting to the Schr\"odinger-like reformulation of the system.

\subsection{Quasi-Normal modes}

Several powerful numerical methods have been developed for the computation of quasinormal modes when the system is of the form (\ref{schroedinger}), but these methods cannot be directly applied to the more general first-order system we are dealing with. In this section, we use a simple numerical method to show  how the Schwarzschild quasinormal modes  can be recovered numerically, using directly the first-order system instead of the Schr\"odinger equation \eqref{eq:schrodinger-like-general}. We restrict ourselves to the polar modes and consider the system (\ref{systemevengen}-\ref{Meven}). The computation of the axial quasinormal modes would  be completely similar. 

By definition of the quasi-normal modes, we impose that the solutions are outgoing at spatial infinity and ingoing at the horizon, which means, using the results of \autoref{subsection_axial}, that the two components of the vector $\X$ satisfy
\bea
\X_1(r) \equiv K(r) & = &  K_\infty(r)  e^{i\omega r_*} = \tilde{K}_\infty(r) e^{i \omega r} r^{i \omega \mass}  \\
       & = &   K_\text{h}(r) e^{-i \omega r_*}  =  \tilde K_\text{h}(r) (r - \mass)^{-i \omega \mass} \, ,
\eea
where $K_\infty$ (and $\tilde{K}_\infty$) is finite at infinity while $K_\text{h}$ (and $\tilde K_\text{h}$) is finite at the horizon, and also
\bea
\X_2(r) \equiv H_1(r)/\omega & = &  H_\infty(r)  r e^{i\omega r_*} = \tilde{H}_\infty(r) e^{i \omega r} r^{1+i \omega \mass}  \\
       & = &   H_\text{h}(r) \varepsilon^{-1} e^{-i \omega r_*}  =  \tilde H_\text{h}(r) (r - \mass)^{-1-i \omega \mass} \, ,
\eea
where again $H_\infty$ (and $\tilde{H}_\infty$) is finite at infinity while $H_\text{h}$ (and $\tilde H_\text{h}$) is finite at the horizon.

Therefore, we look for solutions of (\ref{systemevengen}-\ref{Meven}) using the ansatz
\begin{equation}
    \begin{aligned}
        K(r) = e^{i \omega r} r^{i \omega \mass} \left(\frac{r-\mass}{r}\right)^{-i \omega \mass} f_K(r)\,,\quad
        H_1(r) = e^{i \omega r} r^{1 + i \omega \mass} \left(\frac{r-\mass}{r}\right)^{-1-i \omega \mass} f_H(r) \,,
    \end{aligned}
    \label{eq:ansatz-KR}
\end{equation}
where the functions $f_K$ and $f_H$ are supposed to be finite (hence bounded) both at the horizon and at spatial infinity, in agreement  with the required boundary conditions. Furthermore, we introduce the new variable
\begin{equation}
    u = \frac{2\mass}{r} - 1\,,
\end{equation}
so that the black hole horizon is located at $u = 1$ and spatial infinity at $u = -1$. Each function entering in the equations
\eqref{eq:ansatz-KR} is now treated as a function of $u$ and the system of equations (\ref{systemevengen}-\ref{Meven})
can be expressed  in the form 
\begin{equation}
    \begin{aligned}
    &{\cal P}_{11}(u) f_K(u) + {\cal P}_{12}(u) f_H(u) + {\cal Q}_1(u) f_K'(u) = 0 \,,\\
    &{\cal P}_{21}(u) f_K(u) + {\cal P}_{22}(u) f_H(u) + {\cal Q}_2(u) f_H'(u) = 0 \,,\\
    \end{aligned}
    \label{eq:systeme-diff-u}
\end{equation}
where a prime denotes here a derivative with respect to $u$, and the functions ${\cal P}_{ij}$ and ${\cal Q}_i$ are polynomials in $u$. This is possible because the matrix $M$ given in \eqref{Meven} contains only rational fractions of $r$.

In order to solve the system \eqref{eq:systeme-diff-u} numerically, we adapt the spectral method
presented in \cite{Jansen:2017oag}  and we decompose $f_K(u)$ and $f_H(u)$ onto a basis of Chebyshev polynomials.
The facts that the functions ${\cal P}_{ij}$ and the ${\cal Q}_i$ are polynomials (hence $C^{\infty}$-functions) and the Chebyshev polynomials are bounded at the boundaries ensure
the boundedness of $f_K(u)$ and $f_H(u)$ which is sufficient to enforce the required boundary conditions.  This is called a ``behavioural'' boundary condition~\cite{Boyd1989}.

Then, any  smooth and continuous complex-valued function $g(u)$ defined on the interval 
$[-1,1]$ can be written as an infinite sum of Chebyshev polynomials $T_n(u)$,
\begin{equation}
    g(u) = \sum_{n=0}^\infty g_n T_n(u) \,,
\end{equation}
where $g_n$ are complex coefficients. 
We can approximate the function $g$ by truncating this series at a given order $N$,
the approximation getting better as $N$ is increased. 
Hence, we decompose the two functions $f_K$ and $f_H$ as follows, 
\begin{align}
    f_K(u) \approx \sum_{n=0}^{N} \alpha_n T_n(u) \,,\qquad
    f_H(u) \approx \sum_{n=0}^{N} \beta_n T_n(u) \,,
\end{align}
where $\alpha_n$ and $\beta_n$ are complex coefficients. Notice that the symbol $\approx$ means that we truncated the series at an order $N$, then the equality is not exact.

The next step is to express the differential system \eqref{eq:systeme-diff-u} as a linear system for the coefficients $\alpha_n$ and $\beta_n$,
which is always possible due to fundamental relations satisfied by Chebyshev polynomials\footnote{The Chebyshev polynomials satisfy the properties
\begin{align}
    T'_n(u) =   \sum_{m - n =2k+ 1} \frac{2m}{1+\delta_{n0}}   \, T_m(u) \,, \quad
    (uT_n)(u) = \sum_m \frac12 ((1 + \delta_{n,1}) \delta_{n-1,m} + \delta_{n+1,m})  \, T_m(u) \,,
\end{align}
where $\delta_{m,n}$ is the Kronecker symbol and $k \in \mathbb N$ in the first sum. 
}.  
As a consequence, the differential system \eqref{eq:systeme-diff-u} can be recast as the following system of algebraic equations 
\begin{equation}
    M_{N}(\omega) V_N(\alpha_n, \beta_n) = 0 \,,
\end{equation}
where $M_{N}$ is a $2(N+1) \times 2(N+1)$  matrix whose expansion in powers of $\omega$ reads
\begin{equation}
    M_N(\omega) = M_{N[0]} + M_{N[1]} \omega + M_{N[2]} \omega^2 \,,
\end{equation}
while the $2(N+1)$-dimensional vector  $V_N(\alpha_n,\beta_n)$ is such that
\begin{equation}
    {}^T V_N(\alpha_n, \beta_n) \equiv \begin{pmatrix} \alpha_0, &  \cdots, & \alpha_N, & \beta_0, & \cdots, & \beta_N \end{pmatrix} \, .
\end{equation}
Following \cite{Jansen:2017oag}, we can reformulate this system as
\begin{equation}
    \tilde{M}_{N}(\omega) \tilde{V}_N(\alpha_n,\beta_n) = 0 \,,
    \label{eq:syst-generalized-eigs}
\end{equation}
where the matrix $ \tilde{M}_{N}$  is now of dimension $4(N+1)$ and  defined by
\begin{equation}
    \tilde{M}_N =  \tilde{M}_{N[0]} + \tilde{M}_{N[1]} \omega \quad \text{and} \quad     \tilde{M}_{N[0]} = \begin{pmatrix} M_{N[0]} & M_{N[1]} \\ 0 & I \end{pmatrix} \,, \quad \tilde{M}_{N[1]} = \begin{pmatrix} 0 & M_{N[2]} \\ - I & 0 \end{pmatrix} \,.
\end{equation}

\begin{figure}[!htb]
 \captionsetup{singlelinecheck = false, format= hang, justification=raggedright, font=footnotesize, labelsep=space}
    \includegraphics[scale=1]{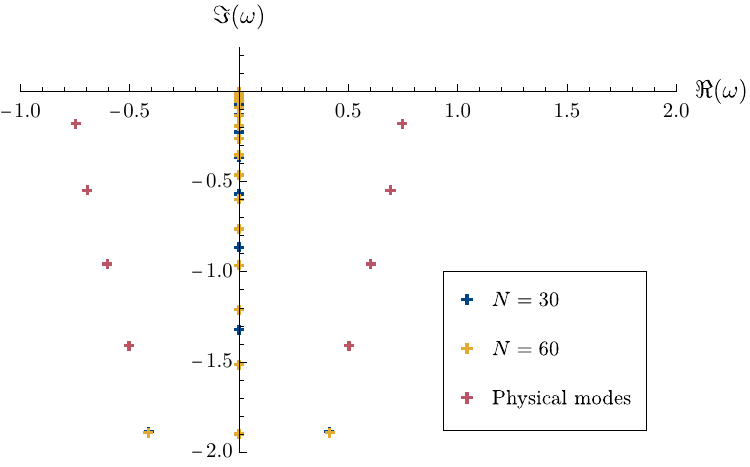}
    \caption{Quasinormal modes numerically found by Mathematica for $\mass = 1$ and $\ell=2$ ($\lambda = 2$). The blue dots are generalised eigenvalues for $N=30$, the orange dots generalised eigenvalues for $N=60$, and the red dots are the modes detected (eigenvalues that change by a factor $10^{-3}$ or less). All the dots present on the imaginary axis correspond to spurious modes. We observe a symmetry with respect to the imaginary axis. The positions of the first modes are $\omega_0 = \pm 0.747 - 0.178i$, $\omega_1 =\pm 0.693 - 0.548i$ and $\omega_2 = \pm 0.602 - 0.957i$. }
    \label{fig:plot-det}
\end{figure}

Finding the values of $\omega$ such that the system \eqref{eq:syst-generalized-eigs} is nontrivial is called a \emph{generalised eigenvalues problem} and can be done by a numerical engine such as Mathematica or Scipy. In practice, we have computed the eigenvalues for different values of $N$ and identified the ones that (almost) coincide when $N$ varies. There are also nonphysical  spurious modes (due to the finite size approximation),  which strongly depend on $N$ and must be discarded. The quasinormal modes thus identified, plotted in Fig.\eqref{fig:plot-det}, coincide with the well-known first quasi-normal modes of Schwarzschild.

This result demonstrates that  it is feasible to 
compute quasinormal modes directly from the first-order system, even if our numerical approach is rather crude and gives a very low precision with respect to the sophisticated methods used in the traditional approach.

\section{General analysis}
\label{sec:general_analysis}
As we have seen in the previous section, it is possible to compute the quasi-normal modes of black holes in general relativity without
reformulating the linearised Einstein equations in terms of  a Schr\"odinger-like equation. The advantage of this method is that it can be straightforwardly generalised to the study of black holes 
 in theories of modified gravity where it might be difficult or impossible to reduce the linearised equations to a  Schr\"odinger-like form. 

In this section, we present a systematic algorithm for a generic first-order system  of the form \eqref{generalsystem}, which has been developed in the mathematics literature, 
 first in \cite{wasow_asymptotic_1965} and more recently in \cite{balser_computation_1999,pflugel_root-free_2019,barkatou_algorithm_1999,barkatou_algorithm_1997,abdelaziz_hilali_solutions_1987}. 
 The various steps of the algorithm presented in this section are summarised in the flowchart diagram depicted in  Appendix \ref{FlowchartApp}. 

\subsection{Asymptotic solution: overview}
We consider a general system of first-order ordinary differential equations of the form
\begin{equation}
  \dv{\X}{z} = M(z) \X \, ,
  \label{eq:ODE}
\end{equation}
where $\X$ is a $n$-dimensional column vector, $M$ an $n\times n$-dimensional  matrix and $z$ a real variable defined in some  
interval. 
In the following, we will consider only the asymptotic behaviour  when  $z \rightarrow +\infty$, but it is straightforward to extend the algorithm near a finite value $z_0$ where the system is singular, by 
a suitable change of the variable $z$.  

We then assume that one can expand $M$ in powers of $z$,
up to some order  (depending
on the required precision of the asymptotic expansion) as follows,
\begin{equation}
\label{generalM}
  M(z)=M_\p z^\p + \dots +M_0 + \dots M_{\p-\q}z^{\p-\q}+ {\cal O}(z^{\p-f-1})= z^\p \sum_{k=0}^{\q} M_{\p-k} z^{-k} + {\cal O}(z^{\p-f-1})\, ,
\end{equation}
where the integer $\p$ is called the {Poincar\'e rank} of the system, and the $M_i$  are $z$-independent matrices. 
In most cases\footnote{Note that, in some cases, the variable $z$  in the expression \eqref{gensol} differs from the variable $z$ in the original system \eqref{eq:ODE}, because a change of variable is necessary, as will be discussed around Eq. \eqref{poverqtrans}. Morever, the special case where $M(z)=M_{-1}/z+ {\cal O}(z^{-2})$ with $M_{-1}$ nilpotent leads to a $\ln z$ behaviour at large $z$, as discussed at the end of  \autoref{Sec:oneblock}.}, the general solution to the system \eqref{eq:ODE} admits an asymptotic expansion of the form \cite{wasow_asymptotic_1965}
\beq
\label{gensol}
\X(z)= e^{\mathbf{\Upsilon}(z)} r^\mathbf{\Delta}\,  \mathbf{F}(z) \X_0\,,
\eeq
where $\X_0$ is a constant vector, corresponding to  $n$ integration constants, $\mathbf{\Upsilon}$ is a diagonal matrix whose coefficients are polynomials of degree at most $\p+1$, $\mathbf{\Delta}$ is a constant diagonal matrix and $\mathbf{F}(z)$ is a matrix which is regular at infinity.

The goal of the algorithm presented below is to determine explicitly the expression (\ref{gensol})
up to some irrelevant sub-leading terms.
As we have already seen in the previous section, the guiding  principle in order to obtain this expression is to fully diagonalise the differential system, up to the appropriate order, by using iteratively transformations of the vector $\X$ into a new vector $\tilde{\X}$,  of the form
\begin{equation*}
	\X(z) = P(z) \tilde{\X}(z) \,,
\end{equation*}
where $P$ is an invertible matrix. The system   \eqref{eq:ODE}  is then transformed into a new but equivalent
differential system, given by
\begin{align}
	\dv{\tilde{\X}}{z} = \tilde{M}(z) \tilde{\X}, \qquad
	\tilde{M}(z) \equiv P^{-1} M P - P^{-1} \dv{P}{z} \,.
	\label{eq:transfo-M}
\end{align}
The end point of this procedure is a system where the matrix coefficients in the expansion of the form \eqref{generalM}  are diagonal at each order. 
It is then immediate to integrate the system and to find the solution in the form \eqref{gensol}, as discussed in  \autoref{sec:Schwarzschild_first_order_method}.

\medskip

In the following subsections, we describe the algorithm step by step. We have also inserted two subsections that contain examples chosen to illustrate some of the finer points of the algorithm. The algorithm contains several branches, depending on whether  the leading term $M_r$ in the expansion of $M(z)$ is diagonalisable or not.

\subsection{Case 1: the leading term is diagonalisable}
\label{sect:diagcase}
The simplest situation is when  the leading matrix $M_r$ is diagonalisable, with each eigenvalue of multiplicity 1.
In this case, one first uses the transformation $\X=P_{(1)} {\X}^{(1)}$ where $P_{(1)}$ is a constant matrix that diagonalises $M_r$, 
which gives the new system 
\bea
	\dv{{\X}^{(1)}}{z} = {M}^{(1)} {\X}^{(1)}, \qquad
	{M}^{(1)}(z) =  D_r z^r +  {M}^{(1)}_{r-1} z^{r-1} + \cdots + {M}^{(1)}_0  + {M}^{(1)}_{-1} \frac{1}{z} + {\cal O}\left( \frac{1}{z^2}\right) \,, 
	\label{diffsystem1}
\eea
where the matrix $D_r$ is diagonal. 

One then seeks to transform the next-to-leading matrix  ${M}^{(1)}_{r-1}$  into a diagonal
matrix (if it is not already) without affecting the diagonal form of the leading order.  This can be accomplished with a new transformation\begin{equation}
\label{P2diag}
\X^{(1)}=P_{(2)} {\X}^{(2)}\,, \qquad	P_{(2)}(z) = \Id + \frac{1}{z} \Xi_{(2)} \,,
\end{equation}
where  $\Xi_{(2)}$ is a constant matrix. Indeed, this yields the new  system
\bea
\dv{{\X}^{(2)}}{z} = {M}^{(2)} {\X}^{(1)}, \quad {M}^{(2)}(z) =  D_r z^r +  D_{r-1} z^{r-1} + {M}^{(2)}_{r-2} z^{r-2}  + \cdots  + {M}^{(2)}_{-1} \frac{1}{z} + {\cal O}\left( \frac{1}{z^2}\right) \,, 
\eea
with
\bea
D_{r-1} \; = \; M_{r-1}^{(1)} + [D_r, \Xi^{(2)}] \, ,
\eea
which is   imposed to be diagonal via an appropriate choice\footnote{To find $\Xi$ such that the matrix
$\tilde D = M + [D ,\Xi]$ is diagonal,   $M$ being arbitrary and $D$  diagonal, one notices that $
[D ,\Xi]_{ij} = (d_i - d_j) \Xi_{ij}$
where $d_i$ are the eigenvalues of $D$. Consequently, $\tilde D$ is given by the diagonal component of $M$ and the coefficients of $\Xi$ satisfy $ (d_i - d_j) \Xi_{ij}  + M_{ij}=0$,  which always admit at least one solution for each $\Xi_{ij}$ as long as all $d_i$ are different.} for  $\Xi_{(2)}$. Furthermore, $D_{r-1}$ is the diagonal part of $M_{r-1}^{(1)}$.

One  can proceed similarly  to ``diagonalise'' all the other terms, order by order, until one gets a system of the form\footnote{Note that we could have proceeded in a single step by  introducing the new variable $\tilde{\X}$ defined by
$\X=P(z)\tilde \X$ with $
P(z) = P_0 + \frac{1}{z} P_1 + \cdots + \frac{1}{z^{r+1}} P_{r+1}$
and determining the constant matrices $P_j$ so that  $\tilde{M}(z)$ is equal to \eqref{algodiag}. The calculation we have just done proves 
this is possible with $\tilde{\X}= \X^{(r+2)}$.} \bea
\label{algodiag}
\dv{{\X}^{(r+2)}}{z} = {M}^{(r+2)} {\X}^{(r+2)}\,, \quad
M^{(r+2)}(z) = D_r z^r + \cdots + D_0 + D_{-1} \frac{1}{z} + {\cal O}\left( \frac{1}{z^2}\right) \,,
\eea
where all matrices are diagonal up to order $1/z$. The system can then be  immediately integrated, to yield
\bea
\X^{{(r+2)}}(z) = e^{\mathbf{\Upsilon}(z)} z^{\mathbf{\Delta}}\,  \mathbf{F}(z) \, \X_0 \, , \quad \mathbf{\Delta} \equiv D_{-1} \, , \quad 
\mathbf{\Upsilon}(z) \equiv D_r \frac{z^{r+1}}{r+1} + \cdots + D_0 z \, ,
\eea
where $\X_0$ is a constant vector.

The asymptotic expansion of the original vector $\X$
can be simply deduced from the combined transformations, i.e.
\bea
\X = P_{(1)} P_{(2)} \cdots P_{(r+2)}   \X^{(r+2)} \, .
\eea
Since the $P_{(j)}$ are polynomials of $1/z$, $\X$ has exactly the same exponential behaviour (in its asymptotic expansion) 
as $\X^{(r+2)}$.

The above procedure is not directly applicable if the leading matrix $M_r$ has eigenvalues of multiplicity higher than one. 
In such a case, writing $M_r$ in a block diagonal form, with eigenvalues $\lambda_i$ of multiplicity $m_i$, one applies a transformation
\begin{equation}
	Y^{(1)} = P_{(2)} Y^{(2)} \,,
\end{equation}
where $P_{(2)}$ has the same block structure as $M_r$,  with the blocks $B_i$ of size $m_i \times m_i$ defined as $B_i = \exp(\frac{\lambda_i}{r+1} z^{r+1}) $ if $m_i \geq 2$  and $B_i=1$   if $m_i = 1$.
For example, if the leading matrix is $M_r={\rm Diag}(\lambda_1, \lambda_1, \lambda_2)$,
with $r = 1$, then the transformation is $P_{(2)} = {\rm Diag}(\exp(\lambda_1 \frac{z^2}{2}),\exp(\lambda_1 \frac{z^2}{2}),1)$.

Such a transformation puts the multi-dimensional blocks to zero, allowing one to pursue the algorithm with the subleading terms. One must however be careful when coming back to the original variable $Y^{(1)}$, since the transformation $P_{(2)}$ will greatly affect the computed asymptotic behaviour.

\subsection{Case 2: the leading term is non-diagonalisable,  similar to a single-block Jordan matrix}

\label{Sec:oneblock}

Solving  asymptotically a system  where the dominant term $M_r$ is not diagonalisable is  more challenging.
The basic idea consists in finding a transformation where  the leading term of the new matrix  becomes diagonalisable.  This can be done by reducing progressively the Poincar\'e rank of the system until the leading term is diagonalisable, in which case the procedure of the previous subsection becomes applicable. If the leading term never gets  diagonalisable
down to  the rank $r=-1$, then the general formula  \eqref{gensol} for the asymptotic 
expansion is not valid but the system  can nevertheless be integrated explicitly. 

 The reduction of the Poincar\'e rank together with the diagonalisation of the leading term is done  in different steps, which we now describe, first when the leading term is similar to a Jordan matrix with a single block. The case of a Jordan matrix with several blocks will be discussed later, in \autoref{section:blockdiag}.

\vspace{0.5cm}
{\it Step 1. Transformation to a Jordan block}
\vspace{0.5cm}

Starting from the asymptotic expansion \eqref{generalM} of the matrix $M$, we use the transformation $X=P_{(1)} X^{(1)}$ to  write $M^{(1)}_r=P_{(1)}^{-1} M_r P_{(1)}$
 in a Jordan canonical form (although with a lower triangular matrix). 
 We assume here that $M^{(1)}_r$ contains   a single (lower triangular) Jordan block with eigenvalue $\lambda$, i.e. of the form
\bea
M^{(1)}_r = 
\begin{pmatrix}
\lambda &0 &  \cdots & &\\
1 & \lambda & 0 &  \cdots &\\
0 & 1 & \lambda & 0 &\cdots \\
\vdots & & &
\end{pmatrix} \equiv \lambda \Id + J(n)\, ,
\eea
where $J(n)$ has the property to be nilpotent (we recall that $n$ is the dimension of the matrix).

\vspace{0.5cm}
{\it Step 2. Transformation to a nilpotent matrix}
\vspace{0.5cm}

We then apply the transformation  
\bea
\X^{(1)}= P_{(2)}{\X}^{(2)}\,, \qquad P_{(2)}(z) \equiv \exp\left( \frac{\lambda}{r+1} z^{r+1} \right) \Id \, ,
\label{eq:exp-shift}
\eea
which renders the leading term nilpotent\footnote{This follows from the relation
\bea
P_{(2)}^{-1} \left(z^r (\lambda \Id +J(n)\right) P_{(2)} - P_{(2)}^{-1}\dv{P_{(2)}}{z} = z^r(M_r^{(1)} - \lambda  \Id)  
= z^r J(n) \,.
\eea
}
\bea
M^{(2)}(z) = J(n) z^r + M^{(2)}_{r-1} z^{r-1} + \cdots + M^{(2)}_0 + M^{(2)}_{-1} \frac{1}{z} + {\cal O}\left( \frac{1}{z^2}\right)  \, .
\label{asymptexpanM2nilp}
\eea

\vspace{0.5cm}
{\it Step 3. Normalisation and reduction of the Poincar\'e rank}
\vspace{0.5cm}

The next step consists in reducing the Poincar\'e  rank of the system by using the transition matrix 
\bea
P(z) = D({n},z) \equiv 
\begin{pmatrix}
1 &0 & 0 & \cdots & \cdots & 0\\
0 & z & 0 &  \cdots &\cdots & 0\\
0 & 0 & z^2 & 0 &\cdots & 0\\
\vdots & & & \ddots & & \vdots \\
0 & \cdots & & & \cdots & z^{n-1}
\end{pmatrix} \, ,
\label{P3nilp}
\eea
which satisfies the useful property
\bea
P^{-1} J(n)P = \frac{1}{z} J(n)  \, .
\eea
A transformation with the above $P$ will thus reduce the order of the leading term $J(n) z^r $, but 
will also affect the sub-dominant terms in the expansion \eqref{asymptexpanM2nilp} of $M^{(2)}$, in particular $M^{(2)}_{r-1}$ which could generate terms whose order is higher than $r-1$ in the new matrix. 

To avoid this situation, we need   first to   ``normalise'' the system, with the transformation
\bea
\label{normalP}
{P}_{(3)}(z)= \Id + \frac{1}{z} \Lambda_{(3)} \, ,
\eea
where $\Lambda_{(3)}$ is a constant matrix, chosen such that such that the next-to-leading order matrix $M^{(3)}_{r-1}$  in the new matrix expansion  contains only zeros  except possibly in the first row. Let us stress that this transformation leaves the leading term of the expansion unchanged. 
The new system associated with ${M}^{(3)}$ is said to be {\it normalised}. 

One can then perform the transformation  generated by the transition matrix
\beq
\label{P_Dn}
P_{(4)}(z) = D({n},z)\,,
\eeq
which, in {\it most} cases, gives a reduced Poincar\'e rank. There are however a few exceptions  where the reduction does not work. These special cases require  a more general transformation, with a transition matrix of the form
\bea
\label{poverqtrans}
P_{(4)}(z) = D({n},z^{p/q}) \, ,\qquad (1\leq p \leq q\leq n)
\eea
where $p$ and $q$ are co-prime integers.
For example, when $n = 4$, the possible choices are  $\{1/4, 1/3, 1/2, 2/3, 3/4, 1\}$, where the last value corresponds to the generic case \eqref{P_Dn}. To identify the appropriate value of $p/q$, one must test successively the possible values, in decreasing order, until the transformation \eqref{poverqtrans} effectively leads to a system with a lower Poincar\'e rank. 
The corresponding value of $p/q$ is said to be ``admissible''. In practice, this can be understood as  a change of variable\footnote{In this case, the asymptotic expansion of the solution may have an exponential behaviour where the argument $Q(z)$ is not a polynomial of $z$ but rather a polynomial of $z^{1/q}$.},  $z$ being replaced by $u = z^{p/q}$. 

\vspace{0.5cm}
{\it Step 4. Diagonalisable or not diagonalisable?}
\vspace{0.5cm}

The next step depends on the nature of the system $(\X^{(4)}, M^{(4)})$, which possesses a lower Poincar\'e rank than the initial system.  If the leading term of $M^{(4)}$ is diagonalisable, one proceeds as in  \autoref{sect:diagcase}.

 If  $M^{(4)}$ is not  diagonalisable,  one needs to reduce again the Poincar\'e rank of the system, unless one has already reached $r=-1$, in which case one can jump directly to the next paragraph. Otherwise, one must distinguish the following different cases. 
\begin{itemize}
\item If the leading term  is similar to a single-block  Jordan matrix and we took $p/q = 1$ in the previous step, we repeat  the procedure of this subsection.
\item If the leading term is similar to a single-block  Jordan matrix but we took $p/q < 1$ in the previous step, we discard the last step, and start again with the normalised system $M^{(3)}$. However, this time, we normalise the system up to second order: after having normalised $M_{-1}$, we repeat the procedure with $z^2$ instead of $z$ in ${P}_{(3)}$ \eqref{normalP} and require that $M_{-2}$ has a specific form. Details can be found in \cite{balser_computation_1999}. If necessary, one can pursue the normalisation to higher orders.
\item If the Jordan canonical form of the leading term contains several blocks, we go to   \autoref{section:blockdiag}.
\end{itemize}

Eventually  we obtain either  a  system with a diagonalisable leading term, which can be solved following  \autoref{sect:diagcase}, or a system of Poincar\'e rank $r=-1$ with
a nilpotent leading term. In the latter case,  the solution
is equivalent to a polynomial of $\ln z$ at large $z$. Indeed, a system of the form
\bea
    \dv{\X}{z} = \frac{\mu_0}{z} \begin{pmatrix}
0 &0 &  \cdots & &\\
1 & 0 & 0 &  \cdots &\\
0 & 1 & 0 & 0 &\cdots \\
\vdots & & &
\end{pmatrix}  \X \, ,
\eea
where $\mu_0$ is an arbitrary constant, is easily integrated.  The components $\X_i$ (for $1 \leq i \leq n$)  are obtained  iteratively and are given by $\X_1(z)=\xi_1$,  $\X_2(z)=\xi_1 \ln z + \xi_2 $ and more generally,
\bea
 \X_i(z) = \sum_{j=1}^i \frac{\xi_j}{(i-j)!} (\mu_0 \ln z)^{i-j} \, ,
\label{lnzbehavior}
\eea
where the $\xi_i$ are $n$ constants of integration. All the components of $\X$ are thus polynomials of $\ln z$ at large $z$.

\subsection{An example with a nilpotent leading term}

Let us give a concrete example of the procedure used for systems with a nilpotent leading term. We  consider the two-dimensional
system defined by
\bea
    \dv{\X}{z} = M(z) \X\,, \qquad
    M(z) = \begin{pmatrix} 0 & 1\\0 & 0\end{pmatrix} z^2 + \begin{pmatrix} 1 & 0\\0 & -1\end{pmatrix} \, ,
  \label{eq:systeme-2}
\eea
and let us determine its asymptotic solution at large $z$, following the algorithm described above.

  We first put the leading term in  its lower triangular Jordan form:
    \begin{equation}
      P_{(1)} = \begin{pmatrix}
      0 & 1 \\
      1 & 0
    \end{pmatrix}
    \quad \implies \quad 
   M^{(1)}(z) = \begin{pmatrix} 0 & 0\\1 & 0\end{pmatrix} z^2 + \begin{pmatrix} -1 & 0 \\ 0 & 1 \end{pmatrix}
  \,.
  \end{equation}
 Since  the leading term is already nilpotent, step 2 is irrelevant. Moreover, the system is  already normalised  since the next-to-leading order term vanishes. 

 We can thus move directly to the reduction of the order of the system and consider the transformation of the form
 \eqref{P3nilp}:
  \begin{equation}
    P_{(2)}(z)  = \begin{pmatrix}
      1 & 0 \\
      0 & z
    \end{pmatrix}
   \quad \implies \quad 
    M^{(2)}(z)  = \begin{pmatrix} 0 & 0\\1 & 0\end{pmatrix} z + \begin{pmatrix} -1 & 0 \\ 0 & 1 \end{pmatrix}\,.
  \end{equation}
  The order has been reduced but the leading term is still nilpotent. Since the reduction was obtained via a transformation with   $p/q = 1$,  we continue the process by doing a new iteration of the algorithm. We first normalise the system with a transformation of the form \eqref{normalP}, 
  \begin{equation}
    P_{(3)}(z)  = I + \frac1z \begin{pmatrix}
      0 & -1 \\
      0 & 0
    \end{pmatrix}
    \quad \implies \quad 
    M^{(3)}(z)  = \begin{pmatrix} 0 & 0\\1 & 0\end{pmatrix} z + \begin{pmatrix} 0 & 1\\0 & -1\end{pmatrix} \frac1z + \begin{pmatrix} 0 & -2\\0 & 0\end{pmatrix} \frac{1}{z^2}\,,
  \end{equation}
 and again reduce the order of the system with
  the transformation
  \begin{equation}
    P_{(4)}(z)  = \begin{pmatrix}
      1 & 0 \\
      0 & z
    \end{pmatrix}
   \quad \implies \quad 
       M^{(4)}(z)  = \begin{pmatrix} 0 & 1\\ 1 & 0\end{pmatrix} + \begin{pmatrix} 0 & -2\\ 0 & -2\end{pmatrix} \frac1z \,.
  \end{equation}
 The leading term is now diagonalisable.  We diagonalise it explicitly, via
  \begin{equation}
    P_{(5)} = \begin{pmatrix}
      -1 & 1\\
      1 & 1
    \end{pmatrix}
    \quad \implies \quad 
    M^{(5)}(z)  = \begin{pmatrix} -1 & 0\\ 0 & 1\end{pmatrix} + \begin{pmatrix} 0 & 0\\ -2 & -2\end{pmatrix} \frac1z\,,
  \end{equation}
 then we diagonalise the next-to-leading term, with a transformation of the form \eqref{P2diag}, 
  \begin{equation}
    P_{(6)}(z)  = \begin{pmatrix}
      1 & 0\\
      1/z & 1
    \end{pmatrix}
     \quad \implies \quad 
    M^{(6)}(z)  = \begin{pmatrix} -1 & 0\\ 0 & 1\end{pmatrix} + \begin{pmatrix} 0 & 0\\ 0 & -2\end{pmatrix} \frac1z + \mathcal{O}\left(\frac{1}{z^2}\right)\,.
  \end{equation}
We have thus managed to fully diagonalise the system, which immediately gives us the asymptotic solution
  \begin{equation}
    \X^{(6)}(z) = \left( 1+ {\cal O}\left({1}/{z} \right) \right)
    \begin{pmatrix}
      \exp(-z) & 0 \\
      0 & \frac{1}{z^2} \exp(z)
  \end{pmatrix} \X_0, \qquad \X_0 \equiv \begin{pmatrix} \xi_1 \\\xi_2 \end{pmatrix} \, ,
  \end{equation}
  where $\X_0$ is a constant column vector. 
 As a consequence, to obtain the behaviour of $\X$ in the original system, we use the combined transformations
  \begin{equation}
    \X =\left(\prod_{j=1}^6 P_{(j)}\right) \X^{(6)},
  \end{equation}
  which implies
  \begin{equation}
    \X(z) = \left( 1+ {\cal O}\left({1}/{z} \right) \right) \begin{pmatrix}
      \xi_1\exp(-z) z^2  + \xi_2\exp(z) \\
      -2 \xi_1 \exp(-z)
  \end{pmatrix}.
  \label{eq:behav-infinity}
  \end{equation}
For this particular example, it turns out that  the original system \eqref{eq:systeme-2} can be solved exactly, with the  solution
\begin{equation}
  \X(z) = \begin{pmatrix} \frac12 \xi_1 \exp(-z) \left(1 + 2z + 2z^2\right) + \xi_2 \exp(z) \\ -2 \xi_1 \exp(-z) \end{pmatrix}.
\end{equation}
One can thus check that the asymptotic solution \eqref{eq:behav-infinity} agrees with the asymptotic behaviour of the exact solution.

\subsection{Case 3: $M_r$ is similar to a  Jordan  matrix with several blocks}
\label{section:blockdiag}
We  now briefly discuss (without entering into too many details, which can be found in \cite{balser_computation_1999}) the more general case where $M_r$ is block diagonalisable and its canonical Jordan form admits several Jordan blocks. The first two steps of  \autoref{Sec:oneblock} still apply to this case and one can find a transformation (with a constant matrix $P$) such that the new  system associated with $M^{(2)}$ (we use the same notation as  in  \autoref{Sec:oneblock}) has a block diagonal leading term $M^{(2)}_r$ with  Jordan lower triangular blocks,  each block being either nilpotent or 1-dimensional:
\bea
M^{(2)}_r  =
\begin{pmatrix}
J(n_1) & 0 &  \cdots & &\\
0 & J(n_2) & 0 &  \cdots &\\
 \vdots & 0 & \ddots & 0 & \cdots\\
 & \vdots & 0 &\lambda_1 & 0 & \cdots\\
 && \vdots & 0 & \lambda_2 & 0\\
 &&& \vdots & 0 & \ddots\\
\end{pmatrix} \, , \qquad
J(n) \equiv
\begin{pmatrix}
0 &0 &  \cdots & &\\
1 & 0 & 0 &  \cdots &\\
0 & 1 & 0 & 0 &\cdots \\
\vdots & & &
\end{pmatrix}.
\label{Mr_block}
\eea
The Jordan form is chosen so that  the blocks $J(n)$ are ordered by decreasing size ($n_1 \geq n_2 \geq \cdots$).
We will use this block structure as a  layout for the block structure of the other matrices  that appear in the expansion of $M^{(2)}$.
And each block will be denoted by two indices,  $(KL)$, corresponding to a submatrix  of dimensions $n_K \times n_L$.

The principle of the diagonalisation procedure is similar to what was done in sections \ref{sect:diagcase} and \ref{Sec:oneblock}. However, it is now possible to have both diagonalisable blocks and nilpotent blocks. Those must be dealt with separately  to get the full asymptotic behaviour  of the system. In order to do this, one can generalise the order-by-order procedure of \autoref{sect:diagcase}: this is called the \enquote{Splitting Lemma} in \cite{balser_computation_1999}. It is not detailed  here, but can be understood by considering blocks instead of scalars in the computations of  \autoref{sect:diagcase}\footnote{In the case where $M^{(2)}_r$ consists of a 2-block Jordan matrix, one would use a transformation of the form
    \begin{equation}
        P = \begin{pmatrix}
            \Id & \sum_{j=1}^{p} \Xi_j \, z^{-j}\\
            \sum_{j=1}^{p}  \Lambda_j \, z^{-j} & \Id
        \end{pmatrix} \, ,
    \end{equation}
    where the $\Xi_j$ and $\Lambda_j$ are constant matrices. Such a transformation, which generalises \eqref{P2diag},
    enables us to transform each $M^{(2)}_{r-j}$  in the same block diagonal form as $M^{(2)}_r$ with a convenient choice of
    $\Xi_i$ and $\Lambda_i$.
    Therefore,  the initial system gives two decoupled sub-systems and, for
    each one, we proceed along the same lines as in the previous section.}.

One can use this lemma to block diagonalise $M^{(2)}$, order by order : the two global blocks considered will be the nilpotent part of $M^{(2)}_r$ and its diagonalisable part. The latter can be dealt with using the procedure given in \autoref{sect:diagcase}, while the former must be addressed using a generalised version of the procedure given in \autoref{Sec:oneblock}. We give here more details about the last part and, in the rest of this section, assume without loss of generality that $M^{(2)}_r$ contains only nilpotent blocks, such that
\begin{equation}
    M^{(2)}_r = \begin{pmatrix}
        J(n_1) & 0 &  \cdots & &\\
        0 & J(n_2) & 0 &  \cdots &\\
        \vdots & 0 & J(n_3) & 0 & \cdots \\
        & \vdots &&&\ddots
    \end{pmatrix}\, \quad ({\rm with}\  n_1 \geq n_2 \geq \cdots \geq n_{\rm last})\,.
\end{equation}
The procedure in such a  case requires to put the system in a specific normalized form. For a  matrix $M$, obtained at a generic step in the algorithm, one says that the matrix is  \enquote{normalized up to order $s$} if all its leading terms ${M}_{r}, \cdots {M}_{r-s}$ have their $(KL)$ blocks verifying the following properties: 
\begin{itemize}
	\item[-] either all rows are equal to zero except possibly the first one if $K \leq L$,
	\item[-] or all columns are  equal to zero except possibly the last one if $K > L$.
\end{itemize}
In order to reach this normalized form, one must use a succession of transformations\footnote{Let us emphasize on the fact that the hierarchy $n_1 \geq n_2 \geq \cdots$ is crucial for this step to succeed.} $P_\text{norm}(k)$ of the form
\begin{equation}
	{P}_{\text{norm}}(k) = \Id + \frac{1}{z^k} \Lambda \,,
	\label{eq:normalisation-blocks}
\end{equation}
where $k$ varies from $1$ to $s$. The matrix $\Lambda$ is a constant matrix, whose coefficients must be chosen, similarly to $\Xi$  in \eqref{P2diag},  such that 
the new matrix $M$ is normalised, in the sense defined above ($\Lambda$ is uniquely defined if one requires that all its blocks $\Lambda^{KL}$ have zero last row if $K \leq L$ and zero first column if $K > L$). The procedure is iterative: if the system is normalized up to order  $k$, it is possible to normalize it up to order $k + 1$ by applying a   transformation $P_\text{norm}(k+1)$. Indeed, this transformation will not modify any term of order higher than $r - k - 1$. 

The complete procedure to reduce the Poincaré rank of the matrix is then the following:
\begin{enumerate}
	\item one starts with $s= 1$ ;
	\item one normalizes the system up to order  $s$ using ${P}_{\text{norm}}(k)$ transformations ;
	\item if ${M}_{r-s}$ is not block-diagonal, one uses a transformation $P_u(n) = \mathrm{diag}(I_{n_1}, I_{n_2}, \cdots, z^s I_{n_{\rm last}})$ and one goes back to step 1;\footnote{It is proved in  \cite{balser_computation_1999} that after a finite number of steps, one always gets a block-diagonal subleading term, which means that this procedure stops at some point and that one can go on with step 4.}
	\item if it is block-diagonal, one uses a $P_{p/q}$ transformation, which is a block form of \eqref{P3nilp} or \eqref{poverqtrans}:
	\begin{equation}
		P_{p/q} = \begin{pmatrix}
			D(n_1,z^{p/q}) & 0 &  \cdots & &\\
			0 & D(n_2,z^{p/q}) & 0 &  \cdots &\\
			\vdots & 0 & D(n_3,z^{p/q}) & 0 & \cdots \\
			& \vdots &&&\ddots
		\end{pmatrix} \, ,
		\label{eq:transfo-block-nilpotent}
	\end{equation}
	where the matrices $D({n},z)$ have been defined in \eqref{P3nilp} and $p$ and $q$ are either co-prime integers (with $1\leq p\leq q\leq n_1$) or equal in the case $p/q=1$ ;
	\item if no $P_{p/q}$ transformation is admissible (see the definition after \ref{poverqtrans}), one goes back to step  1 with $s$ increased by one. Otherwise, one stops here.
\end{enumerate}
Thanks to the above  procedure, one obtains either a system depending on $z$ with a reduced Poincaré rank, or a new system depending on $z^{p/q}$ with a non-nilpotent leading term. In the former case, one can simply pursue with the algorithm. In the latter case, one can change variables by writing $w = z^{p/q}$ and start the algorithm again.

\subsection{A higher dimensional example with $p/q \neq 1$ }
We now present a higher dimensional ($n=5$) example,  adapted from \cite{pflugel_root-free_2019}, where the dominant term in the asymptotic expansion of the matrix $M$ has a non trivial canonical Jordan form with two Jordan blocks. The matrix $M(z)$ is given by
\begin{equation}
  M(z) = \begin{pmatrix}
  0 & z^3 & -z & 1 & 2 z \\
-z^2 & z & 0 & -z & 0 \\
z & 1 & 0 & z^3 & 1 \\
1 & -z & 1 & z & z^3 \\
z & 0 & -3 z & 0 & -1 \\
\end{pmatrix} \equiv  M_3 z^3+M_2 z^2+M_1 z + M_0 \,,
\end{equation}
where the leading term $M_3$ is nilpotent and has a 2-block Jordan structure. 

We perform a first transformation $\X=P_{(1)} \X^{(1)}$  so that the leading term  has now the following  Jordan (lower triangular) canonical form (the matrix $P_{(1)}$ can easily been deduced):
  \begin{equation}
    M^{(1)}(z) = \begin{pmatrix}
    -1 & 0 & -3 z & 0 & z \\
 z^3 & z & 1 & -z & 1 \\
 1 & z^3 & 0 & 1 & z \\
 0 & -z & 0 & z & -z^2 \\
 2 z & 1 & -z & z^3 & 0 \\
\end{pmatrix} \,  \qquad \Longrightarrow \quad
   M^{(1)}_3 = \begin{pmatrix}
    0 & 0 & 0 & 0 & 0 \\
 1 & 0 & 0 & 0 & 0 \\
 0 & 1 & 0 & 0 & 0 \\
 0 & 0 & 0 & 0 & 0 \\
 0 & 0 & 0 & 1 & 0 \\
\end{pmatrix} \, .
  \end{equation}
  The block structure of $M^{(1)}_3$ defines the layout that we will be using to compute the asymptotic expansion of the solution.

  We notice that the next-to-leading term $M^{(1)}_2$ in the  expansion of $M^{(1)}$ is already normalised. Therefore, we can immediately try to reduce the order of the system thanks to a new transformation $\X^{(1)}=P_{(2)} \X^{(2)}$,  
  \begin{equation}
  \label{trans2_cancel}
    P_{(2)} = \begin{pmatrix}
    1 & 0 & 0 & 0 & 0 \\
 0 & z & 0 & 0 & 0 \\
 0 & 0 & z^2 & 0 & 0 \\
 0 & 0 & 0 & 1 & 0 \\
 0 & 0 & 0 & 0 & z \\
    \end{pmatrix}
    \quad\implies\quad 
    M^{(2)} = \begin{pmatrix}
    -1 & 0 & -3 z^3 & 0 & z^2 \\
 z^2 & z-\frac{1}{z} & z & -1 & 1 \\
 \frac{1}{z^2} & z^2 & -\frac{2}{z} & \frac{1}{z^2} & 1 \\
 0 & -z^2 & 0 & z & -z^3 \\
 2 & 1 & -z^2 & z^2 & -\frac{1}{z} \\
    \end{pmatrix} \, .
  \end{equation}
 However, we immediately see that the order of the system has not diminished. This example falls in  the cases where we need to change the variable $z$ or, equivalently, i.e. to make a transformation of the form \eqref{poverqtrans} for each Jordan block, 
 We must therefore cancel the previous transformation \eqref{trans2_cancel} and instead consider  $\X^{(1)}=\tilde{P}_{(2)} \tilde{\X}^{(2)}$, with
 \bea
\tilde{P}_{(2)}(z) = 
\begin{pmatrix}
1 &0 & 0 & 0 &  0\\
0 & z^{p/q} & 0 &  0 & 0\\
0 & 0 & z^{2p/q} & 0 & 0\\
0 & 0& 0& 1 & 0 \\
0 & 0 & 0& 0&  z^{p/q}
\end{pmatrix} \,.
\eea

Following the method described below Eq. \eqref{poverqtrans}, we note that the largest Jordan block is of dimension 3, therefore we should take 2 co-prime integers between \num{1} and \num{3} for $p$ and $q$ with $p \leq q$. The possible choices for the ratio $p/q$ belong
 to the set $\{1/3, 1/2, 2/3\}$, since $p/q=1$ does not work. 
  The largest value is $p/q = 2/3$, which gives for the matrix $M^{(3)}$ the expression
  \begin{equation}
     \begin{pmatrix}
    -1 & 0 & -3 z^{7/3} & 0 & z^{5/3} \\
    z^{7/3} & z-\frac{2}{3 z} & z^{2/3} & -{z}^{1/3} & 1 \\
    \frac{1}{z^{4/3}} & z^{7/3} & -\frac{4}{3 z} & \frac{1}{z^{4/3}} & {z}^{1/3} \\
    0 & -z^{5/3} & 0 & z & -z^{8/3} \\
    2 {z}^{1/3} & 1 & -z^{5/3} & z^{7/3} & -\frac{2}{3 z} \\
  \end{pmatrix}.
  \end{equation}
  We observe that the subdiagonal terms have order $7/3$. To keep this value of $p/q$, we must make sure that no other term behaves like $z^\alpha$ with $\alpha > 7/3$. However in this case there is a $z^{8/3}$ term. Therefore, the value $2/3$ is {not admissible} and we have to consider the next possible choice which is $p/q = 1/2$. Such a change of variable leads to the matrix
  \begin{equation}
    \tilde{M}^{(2)} = \begin{pmatrix}
    -1 & 0 & -3 z^2 & 0 & z^{3/2} \\
    z^{5/2} & z-\frac{1}{2 z} & \sqrt{z} & -\sqrt{z} & 1 \\
    \frac{1}{z} & z^{5/2} & -\frac{1}{z} & \frac{1}{z} & \sqrt{z} \\
    0 & -z^{3/2} & 0 & z & -z^{5/2} \\
    2 \sqrt{z} & 1 & -z^{3/2} & z^{5/2} & -\frac{1}{2 z} \\
  \end{pmatrix}.
  \end{equation}
 Now, it verifies the requirements and  we thus keep the value $p/q=1/2$ and continue the process.

 The previous change of variable leads to a differential system where the coefficients of $M^{(3)}$ are non-integer powers functions of $z$. To apply the algorithm, we have to make a change of coordinate so that the system involves only  integer powers of $z$. This can easily be done by introducing the new coordinate $u$ defined by $z=u^2$. As a consequence, the new differential system is now given by
   \begin{equation}
     \dv{\X^{(3)}}{u} = M^{(3)}(u) \X^{(3)}\, ,\qquad
    M^{(3)}(u) = \begin{pmatrix}
    -2 u & 0 & -6 u^5 & 0 & 2 u^4 \\
2 u^6 & \frac{2 u^4-1}{u} & 2 u^2 & -2 u^2 & 2 u \\
\frac{2}{u} & 2 u^6 & -\frac{2}{u} & \frac{2}{u} & 2 u^2 \\
0 & -2 u^4 & 0 & 2 u^3 & -2 u^6 \\
4 u^2 & 2 u & -2 u^4 & 2 u^6 & -\frac{1}{u} \\
  \end{pmatrix},
  \end{equation}
  where $\X^{(3)}(u) \equiv \tilde{\X}^{(2)}(z)$ and $M^{(3)}(u) \equiv 2u \tilde{M}^{(2)}(z)$ with $z=u^2$.
  As the leading term is not nilpotent, we {keep the value of $p/q$}. If it had been nilpotent, we would have had to go back one step and normalise up to the next order.

 We can continue the algorithm with this new system: we will to do a new change of variables, reduce the order, and decouple the system... We will not present more steps as the rest of the computations is similar to what was done here and in previous sections. Nonetheless, for the sake of completeness, we give the final result. We show that,  after enough steps of the algorithm,  the initial system can be equivalently reformulated as
 \bea
  \dv{\X^{(4)}}{w} = M^{(4)}(w) \X^{(4)}\, , 
 \eea
 where $w = z^{1/6}$ and $M^{(4)}(w)$  is the following diagonal matrix 
  \begin{equation}
      \begin{aligned}
      M^{(4)}(w) = \text{Diag}[
          &3^{4/3} (1 - i \sqrt{3}) w^{19} + 2 w^{11} ,
          -2 \times 3^{4/3} w^{19} + 2 w^{11},\\
          &3^{4/3} (1 + i \sqrt{3}) w^{19} + 2 w^{11} ,
          6 i w^{20} + 3 w^{11},
          - 6 i w^{20} + 3 w^{11}
      ] + \mathcal{O}(w^{9}) \,,
      \end{aligned}
  \end{equation}
up to  order $\mathcal{O}(w^{9})$. Integrating such a system is immediate and yields
the leading orders of the asymptotic expansion of $\X^{(4)}$ from which we can extract the asymptotic expansion of the original variable $\X$.

\section{Conclusion}

In this work, we have studied  the asymptotic behaviours, both at spatial infinity and near the horizon, of the linear perturbations about Schwarzschild black holes. Instead of following the traditional approach that consists in rewriting the equations of motion in the form of a stationary Schr\"odinger-like equation, which is second-order with respect to the radial coordinate, we have worked directly with the first-order equations of motion (in the frequency domain). For this direct approach to the  asymptotic behaviour, we have used an algorithm that has been developed in several recent articles published in mathematical journals. 

The principle of this algorithm is to transform the differential system, via successive changes of  functions, until it can be written in an explicitly diagonal form, up to the required order (in the small parameter characterising the asymptotic regime). This procedure  automatically provides  the combination of the metric perturbations that encapsulates the physical degree of freedom in this asymptotic region and enables one to separate the ingoing and outgoing physical modes.
 Although we have worked in the standard Regge-Wheeler gauge, the same approach would work similarly for any other gauge choice. 

Beyond its application to the perturbations of black holes, this systematic approach to the asymptotic behaviour could be very useful for similar problems in other domains of physics. This is why we have devoted the last part of this paper to a pedagogical presentation of the algorithm, with a few illustrative examples.

For black holes, the knowledge of the asymptotic behaviour of the perturbations is an indispensable first step in the determination of the quasi-normal modes. Indeed, these modes are characterised by the following boundary conditions: a purely outgoing behaviour at spatial infinity and purely ingoing behaviour at the horizon. Imposing these boundary conditions, we have  shown that  the known quasi-normal modes can be recovered numerically,  without resorting to the Schr\"odinger-like formulation, thus providing an alternative approach to the standard method.  We stress that our rudimentary numerical calculation was simply to illustrate the feasibility of this new approach, without trying to reach the precision and efficiency of the powerful numerical methods that have been developed in the traditional approach.

This novel approach could be especially useful in the context of generalised black hole solutions, for instance in modified gravity theories, where the equations of motion for the perturbations are different and extra fields can be present. In a companion paper, we have applied the same algorithm to a few black holes solutions within scalar-tensor theories that belong to the most general known family: DHOST (Degenerate Higher-Order Scalar-Tensor) theories. 
The same method could be applied to the study of other types of black holes, or even completely different physical systems.

As a final remark, let us stress that this approach could be used to get some analytical insight on the asymptotic behaviour of the modes by looking directly at the structure of the matrix coefficients that are relevant. In this sense, it might provide a pre-diagnosis tool to explore the healthiness of some black hole solutions without resorting to a full numerical investigation.

\acknowledgements{
We would like to thank Oleg Lisovyi for instructive discussions and guidance  on the mathematics literature,  as well as Emanuele Berti and Vitor Cardoso for very useful correspondence on quasi-normal modes. We also thank Leo Stein for pointing out a case which was not covered in our previous version.
 We have used Mathematica for many of the calculations involved in this work. KN acknowledges the support from the CNRS grant 80PRIME and thanks the Laboratory of Physics at the ENS in Paris for its hospitality.
}

\appendix

\section{Gauge transformations}
For completeness, we summarise in this Appendix  the gauge fixing procedure for polar and axial  perturbations about a Schwarzschild black hole
in General Relativity, as originally discussed in \cite{Regge:1957td} and \cite{Zerilli:1970se}.

Due to the invariance of the theory under space-time diffeomorphisms,  the metric perturbations are not completely determined $h_{\mu\nu}$. Indeed, any infinitesimal change of coordinates $x^\mu \rightarrow x^\mu + \xi^\mu$  induces the transformation
\bea
\label{diffeo}
h_{\mu\nu} \rightarrow h_{\mu\nu} + \nabla_\mu \xi_\nu + \nabla_\nu \xi_\mu \, 
\eea 
at the linear level. These transformations can be ``projected'' in the axial or polar sectors, which we examine in turn.

\subsection{Axial perturbations: Regge-Wheeler gauge}
\label{Odd-parity perturbations}
Before gauge fixing, axial perturbations are parametrised  by  three families of functions $h_{0}^{\ell m}$, $h_{1}^{\ell m}$ and $h_{2}^{\ell m}$ of the variables $(r,t)$, according to
\bea
&&h_{t\theta} = \frac{1}{\sin\theta}  \sum_{\ell, m} h_0^{\ell m}(t,r) \partial_{\varphi} {Y_{\ell m}}(\theta,\varphi), \qquad
h_{t\varphi} = - \sin\theta  \sum_{\ell, m} h_0^{\ell m}(t,r) \partial_{\theta} {Y_{\ell m}}(\theta,\varphi), \nonumber \\
&&h_{r\theta} =  \frac{1}{\sin\theta}  \sum_{\ell, m} h_1^{\ell m}(t,r)\partial_{\varphi}{Y_{\ell m}}(\theta,\varphi), \qquad
h_{r\varphi} = - \sin\theta \sum_{\ell, m} h_1^{\ell m}(t,r)  \partial_\theta {Y_{\ell m}}(\theta,\varphi), \label{eq:odd-pert} \\
&&h_{ab} =  \sin\theta  \sum_{\ell, m} h_2^{\ell m}(t,r)   \epsilon_{c(a}  D^c \partial_{b)} Y_{\ell m}(\theta,\varphi) \, , \nonumber 
\eea
where, in the last equation, the indices $a$ and $b$ belong to the set $\{\theta,\varphi\}$, $\epsilon_{ab}$ is the totally antisymmetric
symbol such that $\epsilon_{\theta\varphi}=+1$ and $D_a$ is the 2-dimensional covariant derivative associated with the metric
of the 2-sphere $d\theta^2 + \sin^2\theta \, d\varphi^2$. 
{More explicitely, the angular components of the metric can be written
\begin{align}
	h_{\theta\theta} &= \sum_{\ell, m} \frac{1}{\sin\theta} h_2^{\ell m}(t,r) \left(\partial_{\theta}\partial_{\varphi} - \cotan\theta \, \partial_\varphi\right) Y_{\ell m}(\theta,\varphi) \,, \\
	h_{\theta\varphi} &=h_{\varphi\theta} = - \sum_{\ell, m} \sin\theta \, h_2^{\ell m}(t,r) \left( \frac{ \ell(\ell+1)}{2} + \partial_{\theta}^2 \right) Y_{\ell m}(\theta,\varphi) \,, \\
	h_{\varphi\varphi} &= - \sum_{\ell, m} h_2^{\ell m}(t,r) \sin\theta \left(\partial_{\theta}\partial_{\varphi} - \cotan\theta \, \partial_{\varphi} \right) Y_{\ell m}(\theta,\varphi) \,.
\end{align}
All the other components of the axial perturbations vanish. 

In the axial sector, the nonzero components of the generator $\xi^\mu$ that preserves the odd parity of the perturbations can be decomposed into spherical harmonics as follows,
\bea
\xi_\theta = \sum_{\ell,m} \xi^{\ell m}(t,r) \partial_\theta Y_{\ell,m}(\theta,\varphi) \, , \qquad
\xi_\varphi = \sum_{\ell,m} \xi^{\ell m}(t,r) \partial_\varphi Y_{\ell,m}(\theta,\varphi)  \, ,
\eea
and the induced gauge transformations on the functions $h_0$,  $h_1$ and $h_2$  are given, according to \eqref{diffeo}, by
\bea
h_0 \rightarrow h_0 - \dot \xi \, , \qquad
h_1 \rightarrow h_1 - \xi' + \frac{2}{r} \xi \, , \quad h_2 \; \rightarrow \; h_2  - 2 \xi \, ,
\eea
where we have dropped the indices $(\ell m)$ for simplicity. A dot and a prime  denote a derivative with respect to $t$ and $r$,  respectively.

As a consequence, one can always choose a gauge in which $h_2^{\ell m}=0$ which is the well-known Regge-Wheeler gauge for the
axial perturbations \cite{Regge:1957td}. Notice that this gauge choice is possible for $\ell \geq 2$ only (the cases $\ell=0$ and $\ell=1$ will be discussed later below).

\subsection{Even-parity or polar perturbations: Zerilli gauge}
\label{Even-parity perturbations}
Before gauge fixing, polar perturbations of the metric  are
parametrised by seven families of functions $H_{0}^{\ell m}, H_{1}^{\ell m}, H_{2}^{\ell m}$, $\alpha^{\ell m}$, $\beta^{\ell m}$, $K^{\ell m}$ and $G^{\ell m}$ of the variables $(r,t)$ which appear in 
the components of the metric perturbations as follows,
\bea
\label{eq:even-pert}
&&h_{tt} = A(r)\sum_{\ell, m} H_{0}^{\ell m}(t,r) Y_{\ell m}(\theta,\varphi), \quad
h_{tr} = \sum_{\ell, m} H_{1}^{\ell m}(t,r) Y_{\ell m}(\theta,\varphi),\\
&&h_{rr} = \frac{1}{A(r)} \sum_{\ell, m} H_{2}^{\ell m}(t,r) Y_{\ell m}(\theta,\varphi), \\
&& h_{ta} = \sum_{\ell, m} \beta^{\ell m}(t,r) \partial_a Y_{\ell m}(\theta,\varphi), \quad
h_{ra} = \sum_{\ell, m} \alpha^{\ell m}(t,r) \partial_a Y_{\ell m}(\theta,\varphi), \\
&&h_{ab} = \sum_{\ell, m} K^{\ell m}(t,r) g_{ab} Y_{\ell m}(\theta,\varphi) + \sum_{\ell, m} G^{\ell m}(t,r) D_a D_b Y_{\ell m}(\theta,\varphi)\,.
\eea
 {More precisely, the angular part of the metric can be written as
	\begin{align}
		h_{\theta\theta} &=  \sum_{\ell, m} K^{\ell m}(t,r) Y_{\ell m}(\theta,\varphi) + \sum_{\ell, m} G^{\ell m}(t,r) \partial_{\theta}^2Y_{\ell m}(\theta,\varphi)  \,, \\
		h_{\theta\varphi} &=h_{\varphi\theta} = - \sum_{\ell, m} G^{\ell m}(t,r) \cotan\theta \, \partial_\varphi Y_{\ell m}(\theta,\varphi) \,, \\
		h_{\varphi\varphi} &= \sum_{\ell, m} \sin^2\theta \, K^{\ell m}(t,r) Y_{\ell m}(\theta,\varphi) + \sum_{\ell, m} G^{\ell m}(t,r)\left(\partial_{\varphi}^2 + \sin\theta\cos\theta \, \partial_\theta\right)Y_{\ell m}(\theta,\varphi) \,.
	\end{align}

Similarly to the axial sector, this parametrisation is redundant and can be simplified by gauge fixing. Now, linear diffeomorphisms which preserve even-parity of the metric components are generated by
vector fields $\xi$ whose components decompose into spherical harmonics as follows,
\begin{equation}
\begin{aligned}
\xi_t &= \sum_{\ell, m} T^{\ell m}(t,r) Y_{\ell m}(\theta,\varphi) \, , \quad
\xi_r = \sum_{\ell, m} R^{\ell m}(t,r) Y_{\ell m}(\theta,\varphi) \, , \\
\xi_\theta &= \sum_{\ell, m} \Theta^{\ell m}(t,r) \partial_\theta Y_{\ell m}(\theta,\varphi) \, , \quad
\xi_\varphi = \sum_{\ell, m} \Theta^{\ell m}(t,r) \partial_\varphi Y_{\ell m}(\theta,\varphi) \,.
\end{aligned}
\end{equation}
Here $T^{\ell m}$, $R^{\ell m}$ and $\Theta^{\ell m}$ are arbitrary functions of $(t,r)$.
These linear diffeomorphisms induce gauge transformations on the functions that parametrise metric perturbations according to
\bea
\begin{aligned}
H_{0}^{\ell m}(t,r) &\longrightarrow H_{0}^{\ell m}(t,r) + \frac{2 }{A(r)} \dot{T}^{\ell m}(t,r) + A'(r)R^{\ell m}(t,r), \\
H_{1}^{\ell m}(t,r) &\longrightarrow H_{1}^{\ell m}(t,r) + \dot{R}^{\ell m}(t,r) + T'^{\ell m}(t,r) + \frac{A'(r)}{A(r)} T^{\ell m}(t,r), \\
H_{2}^{\ell m}(t,r) &\longrightarrow H_{2}^{\ell m}(t,r) + 2 A(r) R'^{\ell m}(t,r) -A'(r) R_{\ell m}(t,r), \\
\beta^{\ell m}(t,r) &\longrightarrow \beta^{\ell m}(t,r) + T^{\ell m}(t,r) + \dot{\Theta}^{\ell m}(t,r), \\
\alpha^{\ell m}(t,r) &\longrightarrow \alpha^{\ell m}(t,r) + R^{\ell m}(t,r) + \Theta'^{\ell m}(t,r) -\frac{2}{r}\Theta^{\ell m}(t,r), \\
K^{\ell m}(t,r) &\longrightarrow K^{\ell m}(t,r) + \frac{2 A(r)}{r} R^{\ell m}(t,r), \\
G^{\ell m}(t,r) &\longrightarrow G^{\ell m}(t,r) + 2\Theta^{\ell m}(t,r) \, .
\end{aligned}
\eea
An immediate consequence of the gauge transformations is that one can  choose the gauge parameter $\xi$ such that
$G^{\ell m}= 0$ by fixing $\Theta^{\ell m}$, then $\alpha^{\ell m}=0$ and $\beta^{\ell m}= 0$ by fixing $R^{\ell m}$ and $T^{\ell m}$ respectively, in the case where $\ell \geq 2$. This gauge is known as the Zerilli gauge \cite{Zerilli:1970se} (see \cite{Kobayashi:2014wsa} for a recent presentation in the context of modified gravity).

\subsection{Monopole and dipole perturbations}
\label{monopoleanddipole}
We consider here  the special cases $\ell=0$ and $\ell=1$.

\subsubsection{Axial modes}
For the axial modes,  the components $h_{ab}$ vanish identically
for $\ell =1$ (axial perturbations do not have  $\ell=0$ components) which means that $h_2$ does not show up in the components
of the metric. Hence,  when $\ell =1$, it is necessary to make a different gauge choice. 
In general, one chooses $h_1=0$ which fixes the gauge parameter $\xi$ up to a function of the form $C(t) r^2$. Therefore, $h_0$
inherits a residual gauge invariance given by $h_0 \rightarrow h_0 + F(t) r^2 $ where $F(t)$ is an arbitrary function.  Then $h_0$
can be shown to satisfy the equation of motion,
\bea
2 h_0(r) - r h_0'(r) = 0 \, .
\eea 
Therefore, the mode $h_0$ is not propagating.

\subsubsection{Polar modes}
Let us now turn to polar perturbations.
In the case $\ell=0$, $H_0$,
$H_1$, $H_2$ and $K$ are the only non-vanishing components of the metric perturbations whereas $T$ and $R$ are the only
non-vanishing components of the gauge parameter (so that the gauge transformation preserves the monopole). 
As in the general case, one can choose $R$ to fix $K=0$. Then, one can
in principle make use of $T$ to get rid of $H_1$  (we could have also set $H_0=0$). Finally, we are left with only two non-vanishing functions which are either $H_2$ or $H_0$ and we will compute the corresponding equations of motion in the next section.

The main difference, concerning the gauge fixing, between the general case and the case $\ell=1$ lies in the fact that, in the latter, 
$h_{ab}$ can be shown to depend on the difference $G-K$ only, so that one can fix $K=0$ without loss of generality. Furthermore, one 
can make the gauge fixing $G=0$ by an appropriate choice of $\Theta$. Then, one makes use of $T$ to fix $\beta=0$. Finally, one uses
the remaining free gauge function $R$ to fix $\alpha=0$. At the end, we are left with the three non-vanishing functions $H_0$, $H_1$
and $H_2$. The dynamics of these three free parameters will be studied in the next section as well.

\medskip

Concerning the monopole $(\ell=0)$, we showed in  \autoref{Sec:EvenPert} that its dynamics is fully described in terms  of 
the functions $H_0$ and $H_2$ only, as all the
others can be sent to $0$ by gauge fixing. Thus the equations of motion simplifies drastically and, after some calculations, give
\bea
H_0(r) -H_2(r) =0 \, , \qquad H_2(r) + (r-\mass) H_2'(r) = 0 \, .
\eea
The solution reads $H_2(r)=C/(r-\mass)$ and the mode is  not propagating.

\medskip

Finally, the dynamics of the polar dipole $(\ell=1)$ is described by the three non-vanishing functions $H_0$, $H_1$ and $H_2$
which satisfy the three independent equations,
\bea
&&2 H_2(r) + (r-\mass) H_2'(r) = 0 \, , \quad
H_1(r) + i \omega H_2(r) = 0 \, , \nonumber \\
&&H_0(r) +(\mass- r) H_0'(r) - 2 i r \omega H_1(r) + H_2(r)=0 \, .
\eea
Indeed, the full set of the original Einstein equations is equivalent to this one which can easily be solved explicitly but its solution
is not relevant for our purpose. Nonetheless, we see immediately from the equations that, like the monopole, 
the polar dipole does not propagate. This is why we do not consider it  in the rest of the paper.

\section{Equations of motion for the polar perturbations}
\label{App_Eq_even}
In this appendix,  we present the equations of motion satisfied by the  polar perturbations and show how  the system \eqref{eq:systeme-2-eqs} is obtained.
The Euler-Lagrange equations equations of motion \eqref{eom}  yield, in the polar sector,
\begin{equation}
\label{eq:full-eom-schwarzschild}
\begin{aligned}
{\cal E}_{tt} = & - 2(\lambda + 2) \left(1 - \frac{\mass}{r}\right)  H_2(t,r) - 2\lambda \left(1 - \frac{\mass}{r}\right) K(t,r) - \frac{2}{r} (r-\mass)^2 \pdv{H_2}{r}(t,r)\\ &+ \left(6r - 11 \mass + \frac{5 \mass^2}{r} \right)\pdv{K}{r}(t,r) + 2(r-\mass)^2 \pdv[2]{K}{r}(t,r) =0 \, , \\
{\cal E}_{tr} = & -2(\lambda +1) H_1(t,r) - 2r \pdv{H_2}{t}(t,r) + r \frac{2r - 3\mass}{r - \mass} \pdv{K}{t}(t,r) + 2 r^2 \pdv{K}{t}{r}(t,r)=0  \, ,\\
{\cal E}_{rr} =&- 2 \frac{\lambda +1}{1 - \mass/r} H_0(t,r) + \frac{2}{1 - \mass/r} H_2(t,r) + \frac{2 \lambda}{1 - \mass/r} K(t,r) + 2r \pdv{H_0}{r}(t,r) - r \frac{2r-\mass}{2(r-\mass)} \pdv{K}{r}(t,r)\\& - \frac{4r^2}{r-\mass} \pdv{H_1}{t}(t,r) + \frac{2r^4}{(r-\mass)^2} \pdv[2]{K}{t}(t,r) =0 \, , \\
{\cal E}_{t\theta}= & - \frac{\mass}{r} H_1(t,r) - (r-\mass) \pdv{H_1}{r}(t,r) + r \pdv{H_2}{t}(t,r) + r \pdv{K}{t}(t,r) =0 \, ,\\
{\cal E}_{r\theta} = & \frac{2r - 3\mass}{2(r - \mass)} H_0(t,r) - \frac{2r - \mass}{2(r - \mass)} H_2(t,r) - r \pdv{H_0}{r}(t,r) + r \pdv{K}{r}(t,r) + \frac{r^2}{r-\mass} \pdv{H_1}{t}(t,r)=0  \, ,\\
{\cal E}_{\theta\theta} = & \frac{2r+\mass}{2} \pdv{H_0}{r}(t,r) + \frac{2r-\mass}{2}\pdv{H_2}{r}(t,r) - (2r-\mass) \pdv{K}{r}(t,r) + r(r-\mass) \pdv[2]{H_0}{r}(t,r) - r(r-\mass) \pdv[2]{K}{r}(t,r) \\
&- r \frac{2r - \mass}{r - \mass} \pdv{H_1}{t}(t,r) - 2r^2 \pdv{H_1}{t}{r} + \frac{r^3}{r-\mass} \pdv[2]{H_2}{t}(t,r) + \frac{r^3}{r-\mass} \pdv[2]{K}{t}(t,r) =0 \, , \\
{\cal E}_{\theta\varphi} = & H_0(t,r) - H_2(t,r)=0 \, .
\end{aligned}
\end{equation}
 The equations of motion  ${\cal E}_{t\varphi}=0$, ${\cal E}_{r\varphi}=0$ and  ${\cal E}_{\varphi\varphi}=0$ are  identical to ${\cal E}_{t\theta}=0$, ${\cal E}_{r\theta}=0$ and  ${\cal E}_{\theta\theta}=0$, respectively.

We can immediately solve the last equation of the system \eqref{eq:full-eom-schwarzschild} and replace $H_2$ by $H_0$ in all the
other  equations. 
We thus get six equations for only three independent functions  $K$, $H_0$ and $H_1$, { and we want to extract three ``simple'' independent equations out of them}.  
One can then note that  the combination
\begin{equation}
\label{eq:algebraic-equation}
\mathcal{E} \equiv \frac{i \mass}{4 \omega r (r - \mass)} \mathcal{E}_{tr} + \frac12 \mathcal{E}_{rr} + \mathcal{E}_{r\theta} 
\end{equation}
is purely algebraic, i.e.  it does not involve any derivatives of the functions. Moreover, we find that the system $\mathcal{E}_{tr}$, 
$\mathcal{E}_{t\theta}$, $\mathcal{E}_{r\theta}, \mathcal{E}$ enables us  to recover $\mathcal{E}_{tt}$ and $\mathcal{E}_{\theta\theta}$ so that we can  restrict immediately to the system formed by these four equations which, after some simple calculations, are given by
the system of differential equations
\begin{equation}
\begin{aligned}
&K'(r) - \frac{1}{r} H_0(r) - \frac{i (\lambda +1)}{\omega r^2} H_1(r) + \frac{1}{r}\frac{2r-3\mass}{2(r-\mass)} K(r) = 0 \, ,\\
&H_1'(r) + \frac{i \omega r}{r - \mass} H_0(r) + \frac{\mass}{r(r-\mass)} H_1(r) + \frac{i \omega r}{r - \mass} K(r) = 0 \, ,\\
&H_0'(r) - K'(r) + \frac{\mass}{r(r - \mass) } H_0(r) + \frac{i \omega r}{r - \mass} H_1(r) = 0 \, ,\\
\end{aligned}
\label{eq:syst-alg-Schwarzschild}
\end{equation}
{together with the algebraic equation}
\bea
 \left(\frac{3\mass}{r} +2 \lambda \right) H_0(r) + \left( \frac{i \mass (\lambda +1)}{ \omega r^2} - 2 i \omega r\right) H_1(r) + \frac{3\mass^2 + 2\mass(2 \lambda -1)r - 4 \lambda r^2 + 4\omega^2 r^4}{2r (r - \mass)} K(r) = 0 \, . \nonumber
\eea
One equation is still redundant. However, we can solve the  algebraic equation for $H_0$ and substitute its expression into the first three equations. This shows that the third is not independent from the first two. Finally, we obtain
\begin{equation}
\begin{aligned}
K'(r) =&  \frac{3\mass^2 +  \mass( \lambda - 2)r - 2\omega^2 r^4}{r (r-\mass)(3\mass + 2 \lambda r)} K(r)  
+ \frac{i}{\omega r^2} \left(\lambda +1 + \frac{-\mass (\lambda +1) + 2\omega^2 r^3}{3\mass + 2 \lambda r} \right) H_1(r) \, , \\
H_1'(r) =&  \frac{ i r(9 \mass^2 + 8\mass ( \lambda -1)r - 8 \lambda r^2 + 4\omega^2 r^4)}{2(r-\mass)^2 (3\mass + 2 \lambda r)}  \omega K(r)  
- \frac{3\mass^2 + \mass(1+3 \lambda)r - 2 \omega^2 r^4}{r (r-\mass)(3\mass + 2 \lambda r)} H_1(r) \, , 
\end{aligned}
\end{equation}
and  we obtain the required form  \eqref{eq:systeme-2-eqs} with the definitions $X_1(r)\equiv K(r)$ and $X_2(r) \equiv H_1(r) /\omega $.

\section{Flowchart for the algorithm }
\label{FlowchartApp}
{In this appendix, we draw a flowchart to illustrate the algorithm that we are using to compute the asymptotic behaviour of a solution of a first order system.

It should be noted that, in principle, one can skip the first question {\it ``Is the leading term diagonalisable?"} and put directly the leading order term in its Jordan form.  Indeed, when the leading term is diagonalisable, putting it into its Jordan form is equivalent to diagonalising it and the resulting Jordan matrix is made of $d$ one-dimensional blocks where $d$ is the dimension of the system, thus of the matrix. Therefore, the procedure for splitting the system into several subsystems described in  \autoref{section:blockdiag} is in this case equivalent to the procedure described in  \autoref{sect:diagcase} where we are treating several blocks.}

\begin{figure}[h!]
 \captionsetup{singlelinecheck = false, format= hang, justification=raggedright, font=footnotesize, labelsep=space}
\label{flowchart}
    \centering
\begin{tikzpicture}
    \matrix[
    matrix of nodes,
    column sep =20pt,
    row sep = 20pt,
    ]{
        |[name=start, startstop]|   {\textbf{Start}:\\ Consider the leading term in the expansion
        of $M(z)$}\\
        |[name=stop-cond, decision]| {Is the leading term diagonalisable?}
            &|[name=diag-step, startstop]| {Diagonalise step by step with transformations in  \autoref{sect:diagcase} and \textbf{Stop}} \\
        |[name=Jordan, process]| {Put it under the form \eqref{Mr_block}} \\
        |[name=sev-eigenvals, decision]| {Does it have diagonalisable blocks?}
            &|[name=splittinglemma, process, text width=5cm, minimum width=5cm]| {Split the system into its nilpotent and diagonalisable part (Splitting Lemma of \ref{section:blockdiag}) and proceed with each part separately}\\
        |[name=rminusone, decision]| {Is the Poincaré rank $r$ equal to -1?} & &|[name=stopminusone, startstop]| {\textbf{Stop}: compute the logarithmic asymptotic behaviour (see \ref{lnzbehavior})} \\
        |[name=normalize, process]| {Follow the steps of \autoref{section:blockdiag} to get a matrix with block-diagonal subleading order} & &|[name=normalize-next, process]| {Normalise the next order} \\
        |[name=reducing, process]| {Use a transformation of the form \eqref{P3nilp} for each block (\eqref{eq:transfo-block-nilpotent} with $p/q = 1$) to try to reduce the order} & & \\
        |[name=is-reduced, decision]| {Is the order reduced?} &|[name=p-q, process]| {Try again with a change of variables $z \rightarrow z^{p/q}$ (see Eq. \ref{eq:transfo-block-nilpotent})} &|[name=still-nilpotent, decision]| {Is the leading order still nilpotent?}\\
    };

    \draw [arrow] (start) -- (stop-cond);
    \draw [arrow] (stop-cond) -- node[anchor=east] {no} (Jordan);
    \draw [arrow] (stop-cond) -- node[anchor=south] {yes} (diag-step);
    \draw [arrow] (Jordan) -- (sev-eigenvals);
    \draw [arrow] (sev-eigenvals) -- node[anchor=south] {yes} (splittinglemma);
    \draw [arrow] (sev-eigenvals) -- node[anchor=east] {no} node[align=center, anchor=west] {(all blocks\\are nilpotent)} (rminusone);
    \draw [arrow] (splittinglemma.east) -- node {} ++(1., 0) |- ([yshift=-0.5cm]start.east);
    \draw [arrow] (rminusone.east) -- node[anchor=south] {yes} (stopminusone.west);
    \draw [arrow] (rminusone) -- node[anchor=east] {no} (normalize);
    \draw [arrow] (normalize) -- (reducing);
    \draw [arrow] (reducing) -- (is-reduced);
    \draw [arrow] (is-reduced) -- node[anchor=south] {yes} ++(-3.5,0) |- (start);
    \draw [arrow] (is-reduced) -- node[anchor=south] {no} (p-q);
    \draw [arrow] (p-q) -- (still-nilpotent);
    \draw [arrow] (still-nilpotent) -- node[anchor=east] {yes} (normalize-next);
    \draw [arrow] (normalize-next) -- (normalize);
    \draw [arrow] (still-nilpotent) -- node[anchor=south] {no} ++(3.5, 0) |- ([yshift=0.5cm]start.east);
\end{tikzpicture}
\end{figure}

\newpage

\bibliographystyle{utphys}
\bibliography{biblio_QNM}

\end{document}